\documentclass[preprint,11pt]{elsarticle}

\usepackage[english]{babel}

\usepackage[letterpaper,top=2cm,bottom=2cm,left=3cm,right=3cm,marginparwidth=1.75cm]{geometry}

\usepackage{amsmath}
\usepackage{amssymb}
\usepackage{graphicx}
\usepackage{subcaption}
\usepackage[colorlinks=true, allcolors=blue]{hyperref}
\usepackage{adjustbox}
\usepackage{float}
\usepackage{tikz}
\usetikzlibrary{arrows.meta}
\usepackage{caption}
\usepackage{svg}
\usepackage{soul}
\usepackage{xcolor}
\usepackage{comment}
\usepackage{units}
\usepackage{ulem}
\usepackage{lineno}
\DeclareMathAccent{\svec}{\mathord}{letters}{126}
\newcommand\stvec[1]{\mathbf #1}				
				
\newcommand\ssvec[1]{\svec{\stvec{#1}}}	
	
\newcommand\cssvec[1]{\svec{\tilde{\stvec{#1}}}}

\journal{Journal of Wind Engineering and Industrial Aerodynamics}

\begin{document}

\begin{frontmatter}
\title{Modelling Wind Turbines via Actuator Lines in High-Order $h/p$ Solvers}

\author[1]{Oscar A. Marino}
\cortext[cor1]{Corresponding author}
\ead{oscar.marino@upm.es}
\author[1]{Raúl Sanz}
\author[1]{Stefano Colombo}
\author[1]{Ananth Sivaramakrishnan}
\author[1,2]{Esteban Ferrer}

\address[1]{ETSIAE-UPM-School of Aeronautics, Universidad Politécnica de Madrid, Plaza Cardenal Cisneros 3, E-28040 Madrid, Spain}
\address[2]{Center for Computational Simulation, Universidad Politécnica de Madrid, Campus de Montegancedo, Boadilla del Monte, 28660 Madrid, Spain}


\begin{abstract}

This paper compares two actuator line methodologies for modelling wind turbines employing high-order $h/p$ solvers and large-eddy simulations. The methods combine the accuracy of high-order solvers (in this work the maximum order is 6) with the computational efficiency of actuator lines to capture the aerodynamic effects of wind turbine blades. 
Comparisons with experiments validate the actuator line methodologies. We explore the effects of the polynomial order and the smoothing parameter associated with the Gaussian regularization function, and use them to blend the actuator line forcing in the high-order computational mesh, to show that both parameters influence the distribution of forces along the blades and the turbine wake. The greatest impact is obtained when the polynomial order is increased, allowing one to better capture the physics without requiring new meshes.

When comparing the actuator line methodologies, we show the advantages of performing weighted sums, over element averages, to compute the blade velocities and forces in high-order solvers. For low-order simulations (low polynomial orders), both methods provide similar results, but as the polynomial order is increased, the weighted sum method shows smoother thrust/torque distributions and better wake resolution. Furthermore, cell averaging introduces nonphysical oscillations when increasing the polynomial order beyond 3 (4th order accuracy).

\end{abstract}

\end{frontmatter}

\section{Introduction}
Accurate modelling of wind turbine flows is crucial to understanding their performance and optimizing their design. Numerical simulations play a vital role in this process, providing valuable insights into the flow physics around wind turbine blades and associated turbine wakes. 

Large Eddy Simulations (LES) are often used to simulate atmospheric turbulent flows where turbines are represented using actuator discs or actuator lines \cite{Metha,Breton}, to reduce the computational cost of simulating detailed blades. 
Indeed, to investigate the behavior of wakes and their interaction, resolving the near blade boundary layer is of secondary importance and the computational cost can be alleviated by modelling the blades (and avoiding to resolve the flow around them using, for example, sliding meshes). Simulations using actuator discs and actuator lines have been very useful in providing insight into the optimization of wind turbine farm layouts; see \cite{doi:10.1146/annurev-fluid-010816-060206}. 

The pioneering work by Sørensen and Shen \cite{Sorensen_2002, Shen_2010, sorensen2015simulation} laid the foundation for the application of actuator line (AL) methods in the context of wind energy. This method represents the rotating turbine blades as lines of distributed forces, known as actuator lines. The AL mimics the influence of the blades on the flow field, allowing for a computationally affordable representation of the rotor blades but capturing wakes accurately, for example, \cite{Troldborg_2008}. 
Validation of actuator line methods against experimental wind tunnel measurements and full-scale field data has shown that these techniques can provide accurate solutions at limited costs \cite{Churchfield_2012, Sorensen_2015}. Various turbulence models, such as LES and Reynolds-averaged Navier–Stokes (RANS), have been integrated with ALs to account for turbulence effects on both the wake structure and turbine performance \cite{Yang_2015, Kang_2019}. 
Integrating ALs with LES allows for an accurate representation of time-varying and turbulent effects, capturing intricate wake dynamics and tip vortices \cite{Troldborg_2008, Sorensen_2015}. AL predictions for wake flow have been extensively validated against wind tunnel measurements in both single-wind turbine and wind farm scenarios, demonstrating good agreement in both the near and far wake regions \cite{Wu_PorteAgel_2011, sorensen2015simulation, Stevens_2018, Doubrawa_2020, Draper_2018, Uchida_2022}.
When comparing ALs and blade-resolved simulations under non-sheared inflow conditions, Troldborg et al. \cite{Troldborg_2015} found that the actuator disk and AL produced similar wakes to resolved blades when the inflow was turbulent. However, under laminar ambient conditions, resolving the blade boundary layer resulted in higher turbulence and eddy viscosity in the near wake, leading to significant differences in the wake several diameters downstream \cite{Troldborg_2015}.
Recent studies have focused on improving the accuracy of the aerodynamic polars (lift and drag coefficients) employed in the simulation of ALs \cite{Mikkelsen_2003, Martnez_Tossas_2016}. Additionally, efforts have been made to incorporate unsteady aerodynamics into the modelling process, capturing dynamic stall and other time-dependent effects \cite{Lignarolo_2019}. Recently, \cite{ARABGOLARCHEH2022871,Arabgolarcheh_2023} employed the AL technique to simulate wind turbines in offshore environments. Finally, there is a growing interest in extending ALs to underwater tidal turbines, see for example \cite{FREDRIKSSON20211140,BABAAHMADI2017420,LIU2019473,BABAAHMADI2017669}.

In recent years, there has been a growing interest in incorporating AL models into high-order $h/p$ solvers. 
High order $h/p$ methods are very flexible and can improve the numerical accuracy by either increasing the number of elements in the mesh ($h$-refinement) or by increasing the polynomial order in each mesh element ($p$-refinement). The former is the only way to increase the accuracy in low order methods (finite volumes, finite elements, etc.) while the latter is only available in high order $h/p$ solvers. High-order polynomials produce an exponential decay of the error for smooth solutions instead of the algebraic decay characteristic of low-order techniques. Overall, increasing the polynomial order leads to a more efficient use of the degrees of freedom, see for example \cite{HORSES_3D} and references therein.
In high-order discretizations, the formal order of the scheme is $p+1$, where $p$ is the polynomial order within each mesh element. 
In this work, we will use polynomial orders ranging from $p=2$ to 5 (i.e., 3rd to 6th order). 

By combining the actuator line method with high-order $h/p$ solvers, researchers aim to enhance the fidelity of wind turbine simulations and provide more accurate predictions of turbine performance metrics, including power production and aerodynamic loads. In the high-order context (continuous formulations) a series of papers have adapted AL to $h/p$ meshes \cite{https://doi.org/10.1002/we.2253,Datos_numericos,https://doi.org/10.1002/we.2458} for a variety of turbulence models ranging from unsteady RANS to LES (modeled through a spectral vanishing viscosity models). Key questions on how to implement actuator lines in high-order methods remain and are the topic of this work. Namely, we study the effect of the polynomial order, the Gaussian smoothing parameter, and two ways to compute the forces on the actuator lines: averaging the velocity in each cell or using a weighted average forcing.

We investigate the application of actuator lines within the high-order discontinuous Galerkin $h/p$ solver and focus on the numerical implementation and validation of the method against experimental data. The objective is to assess the accuracy, computational efficiency, and robustness of the high-order actuator lines. Through comparisons with experimental data, we demonstrate the capabilities of the AL method in capturing complex flow phenomena around wind turbine blades and provide guidelines for the implementation of AL in $h/p$ solvers to simulate wind turbines.

The rest of the paper is organized as follows. Section \ref{sec:meth} presents the high order solver and the two actuator line methodologies considered in this work. The turbine and numerical parameters selected for the study are detailed in Section \ref{sec:definition}. Section \ref{sec:results} summarises the results and discussion. Finally, conclusions are included in are in Section \ref{sec:conclusion}.

\section{Methodology}\label{sec:meth}

\subsection{High order discontinuous Galerkin $h/p$ solver (HORSES3D)}

HORSES3D \cite{HORSES_3D} is an open source $h/p$ solver (\textit{https://github.com/loganoz/horses3d}) that discretizes the Navier-Stokes equations using the Discontinuous Galerkin Spectral Element Method (DGSEM), which is a highly efficient nodal version of discontinuous Galerkin (DG) family of methods \cite{2009:Kopriva}. 
One of the main advantages of DG methods is their ability to accurately capture high-order spatial and temporal variations of the solution, which makes them particularly suitable for simulating flows with sharp gradients and complex flow phenomena. DG methods also exhibit good numerical stability and conservation properties because of the local nature of the approximation and the use of fluxes at element interfaces. 

 In this section, we provide only a brief overview of the fundamental concepts of DG discretizations for the compressible Navier-Stokes retained in this work; see the Appendix \ref{sec:cNS}.
The physical domain is subdivided into nonoverlapping curvilinear hexahedral elements, $e$, which are geometrically transformed to a reference element, $el$, using a polynomial transfinite mapping that relates the physical coordinates $\vec{x}$ and the local reference coordinates $\vec{\xi}$. This transformation is applied to Eq.~\eqref{eq:compressibleNScompact_transformed}, resulting in:

\begin{equation}
\boldsymbol{q}_t  + \nabla_x\cdot\ssvec{F}_e = \nabla_x\cdot\ssvec{F}_{v,turb}+\boldsymbol{S({q}_t)},
\rightarrow
J \boldsymbol{q}_t  + \nabla_\xi\cdot\cssvec{F}_e = \nabla_\xi\cdot\cssvec{F}_{v,turb}+J \boldsymbol{S({q}_t)},
\label{eq:compressibleNScompact_transformed}
\end{equation}
where $J$ is the Jacobian of the transfinite mapping, $\nabla_\xi$ is the differential operator in the reference space and $\cssvec{F}$ are the contravariant fluxes \cite{2009:Kopriva}. 
To derive DG schemes, we multiply Eq.~\eqref{eq:compressibleNScompact_transformed} by a locally smooth test function $\phi_j$ ($0\leq j\leq P$, where $p$ is the polynomial degree) and integrate over an element $el$ to obtain the weak form:

\begin{equation}\label{eq::NS2}
\int_{el}J \boldsymbol{q}_t\phi_j+\int_{el} \nabla_\xi\cdot\cssvec{F}_e\phi_j  =\int_{el} \nabla_\xi\cdot\cssvec{F}_{v,turb}\phi_j+\int_{el}J\boldsymbol{S({q}_t)}\phi_j.
\end{equation}

By integrating the term with the inviscid fluxes, $\mathbf{F}_e$, by parts, we obtain a local weak form of the equations (one per mesh element) with the boundary fluxes separated from the interior:
\begin{equation}\label{eq::NS3}
\int_{el}J \boldsymbol{q}_t\phi_j +  \int_{\partial el} \cssvec{F}_e\cdot\hat{\mathbf{n}}\phi_j-\int_{el} \cssvec{F}_e\cdot\nabla_\xi\phi_j
=\int_{el} \nabla_\xi\cdot\cssvec{F}_{v,turb}\phi_j+\int_{el}J\boldsymbol{S({q}_t)}\phi_j,
\end{equation}
where $\hat{\mathbf{n}}$ is the unit outward vector of each face of the reference element ${\partial el}$.
Discontinuous fluxes at inter-element faces are replaced by a numerical inviscid flux $\mathbf{F}_{e}^{\star}$, to couple the elements:
\begin{equation}\label{eq::NS4}
\int_{el}J \boldsymbol{q}_t\cdot\phi_j + \int_{\partial el} {\cssvec{F}_{e}^{\star}}\cdot\hat{\mathbf{n}}\phi_j-\int_{el} \cssvec{F}_e\cdot\nabla_\xi\phi_j
=\int_{el} \nabla_\xi\cdot\cssvec{F}_{v,turb}\phi_j+\int_{el}J\boldsymbol{S({q}_t)}\phi_j.
\end{equation}

The equations for each element are coupled through Riemann fluxes $\mathbf{F}_{e}^{\star}$. Viscous terms are also integrated by parts but require further manipulations to obtain usable discretizations \cite{unified,HORSES_3D}. The nonlinear inviscid and viscous (including turbulent) numerical fluxes can be chosen appropriately to control dissipation in the numerical scheme~\cite{Manzanero_2020,Ferrer_2017,jumpKou}. In this work, we use the Lax-Friederich flux and Bassi-Rebay 1 (BR1) schemes, respectively. The volume form is given by:

\begin{equation}\label{eq::NS5}
\int_{el}J \boldsymbol{q}_t\cdot\phi_j + \underbrace{{ \int_{\partial el} \cssvec{F}_{e}^{\star}}\cdot\hat{\mathbf{n} \phi_j}}_\text{Convective fluxes}-\int_{el} \cssvec{F}_e\cdot\nabla_\xi\phi_j
= \underbrace {\int_{el} (\nabla_\xi\cdot\cssvec{F}_{v,turb})\cdot\phi_j }_\text{Viscous and Turbulent fluxes} +\underbrace{ \int_{el}J\boldsymbol{S({q}_t)}\phi_j}_\text{Source term for actuator line}.
\end{equation}

In a final step, we approximate the numerical solution and fluxes using polynomials (of order $p$) and evaluate all integrals using Gaussian quadrature rules. In this work we use Gauss-Legendre points for the latter. In addition, we have complemented the compressible Navier-Stokes equations with the Vreman \cite{Vreman_2004} Large Eddy Simulation subgrid model. Note that the source term $\mathbf{S}$ will be used to incorporate the actuator line forcing. Finally, to advance the solution in time we use a Runge-Kutta 3 explicit time marching scheme (with a $CFL<1$ in all cases).

\subsection{Two actuator line methodologies}
Actuator lines \cite{sorensen2015simulation} are popular methods to simulate wind turbines using low-order methods (order 2, equivalent to $p=1$) such as finite volumes. In this work, we reinterpret classic actuator line implementations from the perspective of high-order ($p\geq2$, order larger than 3) discretizations. 

ALs model the wind turbine blades via a line of points (actuator points) that represent the aerodynamic centers of the airfoil sections defining the blade. To incorporate the forces of the sectional blade, the AL employs blade element theory, which calculates the aerodynamic forces and moments on each line element based on its local angle of attack, airfoil aerodynamics (lift and drag coefficients) and inflow conditions. These forces are then distributed to the surrounding flow. 

To calculate the forces, tabulated lift and drag coefficients for different angles of attack and Reynolds numbers are used:
\begin{equation}
f_L = \frac{1}{2} \rho U_{rel}^2 S C_l, \
\quad f_D = \frac{1}{2}\rho U_{rel}^2 S C_d, \
\label{eq: FL FD}
\end{equation}
where $S$ is the sectional area, $\rho U_{rel}^2$ is the local dynamic pressure and $C_l$,$C_d$ are the sectional aerodynamic coefficients (e.g., 2D forces calculated from XFOIL). In this study, we fix the number of blade sections for all formulations and parameters. 

The AL requires sampling the flow velocity near the blade axis to calculate the forcing. In this work, we sample the velocity at the location where the forcing is applied (classic technique), which corresponds to a Gauss-Legendre point inside a mesh element. High-order order discretization considers a polynomial inside each element and thus sampling a point takes into account the whole polynomial inside the mesh elements and the parasitic effect of sampling where we applied the force is reduced, and we observe good numerical behavior. Let us note that other methods have been proposed to sample the velocity near the application point; see, for example \cite{Schluntz_2015}), including filtering processes~\cite{Stanly_2022}.

\begin{figure}[H]
    \begin{subfigure}{.52\textwidth}
        \begin{tikzpicture}
            \draw (0.0,0) node[inner sep=0]{ \includegraphics[trim={4cm 5.8cm 4cm 3.6cm}, clip,width=0.98\linewidth]{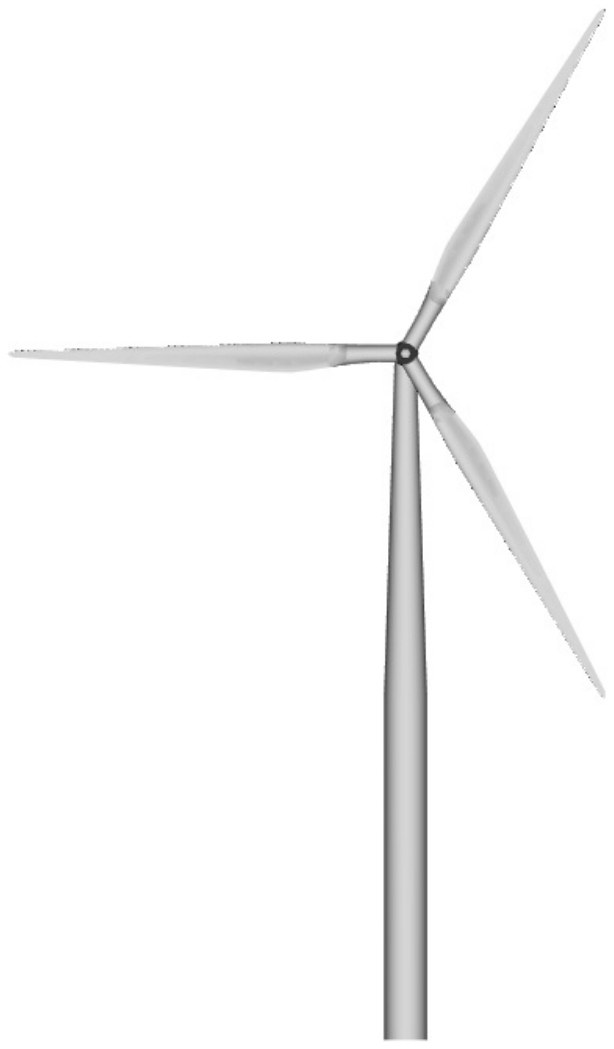} };
            \draw [thick] (0.33,1.1) arc (180:40:0.5cm) node[above,xshift=-18,yshift=8]{$\theta$};
            \draw [ultra thick,arrows=->] (1.0,1.1) -- (-4.2,1.1) node[below left]{$y$};
            \draw [ultra thick,arrows=->] (1.0,1.1) -- (1.0,5.8) node[above left]{$z$};
            \draw [dashed] (1.0,1.1) -- (3.64,5.62);
            \draw[ultra thick, arrows=->] (-1.5,2) arc (180:60:1) node[below left,xshift=-5]{$\Omega$};
        \end{tikzpicture}
	\end{subfigure}
    \begin{subfigure}{.37\textwidth}
        \begin{tikzpicture}
            \draw (0.0,0) node[inner sep=0]{ \includegraphics[trim={7cm 6cm 6cm 4cm},clip,width=0.85\linewidth]{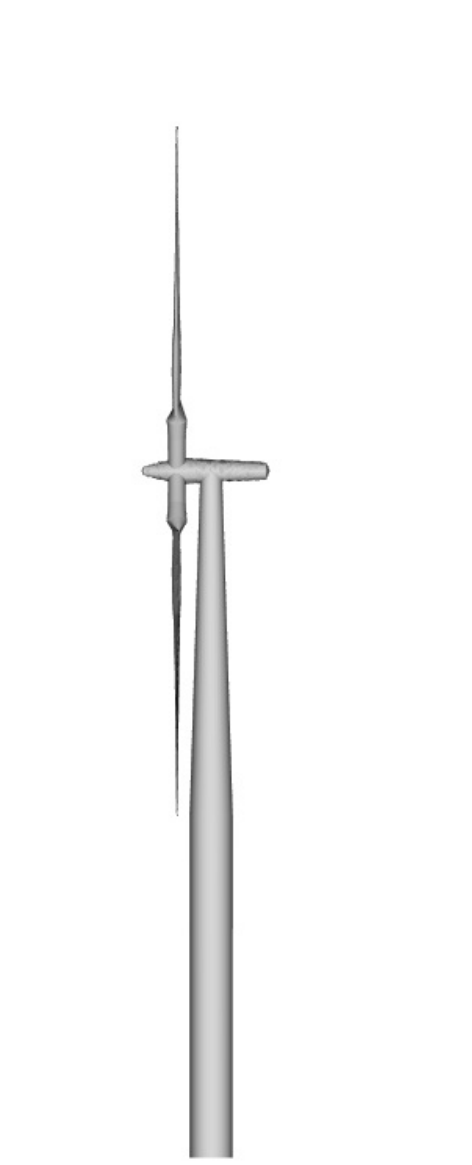} };
            \draw [ultra thick,arrows=->] (-0.63,1.1) -- (3.2,1.1) node[below right]{$x$};
            \draw [ultra thick,arrows=->] (-0.63,1.1) -- (-0.63,5.8) node[above left]{$z$};
        \end{tikzpicture}
	\end{subfigure}
    \centering
    \caption{Wind turbine reference frame.}
    \label{fig:frame}
\end{figure}

Consider the global inertial axes, $[x, y, z]^T$, where $x$ represents the direction of incident flow, $y$ is perpendicular to the free-stream velocity, and $z$ is the local vertical axis, as depicted in Figure~\ref{fig:frame}. We follow \cite{Liu_2022} to summarize the fluid kinematics around the blade sections.

Taking into account the instantaneous flow velocity at a generic location as $\mathbf{v} = [v_x, v_y, v_z]^T$, it can be converted to cylindrical coordinates using the reference frame illustrated in \ref{fig:frame}. At any actuator point, the azimuthal component of the local velocity, $U_\theta$, can be calculated as:
\begin{equation}
U_\theta = \Omega r + v_y \sin{\theta} - v_z \cos{\theta},
\label{eq: u_theta}
\end{equation}

\noindent where $\Omega$ is the rotational speed of the actuator line, $r$ is the radius of the actuator point, and $\theta$ is the azimuthal angle. Similarly, the angle of attack for each section is defined as
\begin{equation}
\alpha = \phi - \gamma; \quad \phi = \arctan\left(\frac{U_x}{U_\theta}\right),
\label{eq: angle of attack and local}
\end{equation}
\noindent where $\phi$ is the local angle of incidence, $\gamma$ is the angle due to local twist and pitch of the airfoil, and $U_x=v_x$ is the axial component of the local velocity.

The relative velocity, $U_{rel} = \sqrt{U_x^2 + U_\theta^2}$, is obtained, and the forces $\stvec{f} = [f_x,f_y,f_z]^T$ are projected onto the global axes following:
\begin{equation}
\begin{split}
&f_x = - (f_L\cos(\phi) + f_D\sin(\phi)),\\
&f_y = (f_L\sin(\phi) - f_D\cos(\phi))\sin(\theta),\\
&f_z = - (f_L\sin(\phi) - f_D\cos(\phi))\cos(\theta).\\
\end{split}
\label{eq: Fx Fy Fz}
\end{equation}

To avoid introducing point-wise forces into the solver, which can create numerical instabilities, each of these forces is smeared across various computational nodes (in our case various Gauss-Legendre nodes inside one mesh element). This is achieved by using a Gaussian regularization expression:

\begin{equation}
    \eta_{\epsilon} (d) = \frac{1}{\epsilon^3 \pi^\frac{3}{2}} e ^{-(\frac{d}{\epsilon})^2}
    \label{eq: eta_eps},
\end{equation}

\noindent where:
\begin{equation}
    d = \sqrt{(x_{i,j,k} - x_a)^2 + (y_{i,j,k} - y_a)^2 + (z_{i,j,k} - z_a)^2},   
    \label{eq: distance}
\end{equation}
is the distance between an actuator point and an arbitrary node. The subscript $a$ is used to indicate the actuator forcing point, and the subscripts $i$, $j$, and $k$ represent a node. Thus, the source term used to smooth the actuator line forcing in Eq.~\ref{eq::NS5} is:

\begin{equation}
    \mathbf{S(q_t)} = \eta_{\epsilon} \mathbf{f}.
    \label{eq: source_term_eqn}
\end{equation}

\begin{figure}[H]
    \centering
\begin{subfigure}{0.4\textwidth}
    \includegraphics[scale = 0.4]{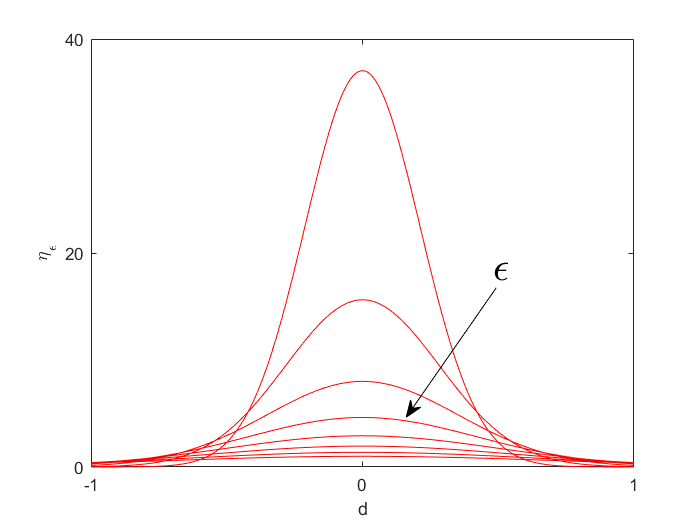}
    \captionsetup{justification=centering}
    \caption{Gaussian projection function for different values of $\epsilon_k$. $d$ denotes the local support distance for the Gaussian.}
    \label{fig: eta_eps}
\end{subfigure}
\hfill
\begin{subfigure}{0.55\textwidth}
    \includegraphics[scale=0.4]{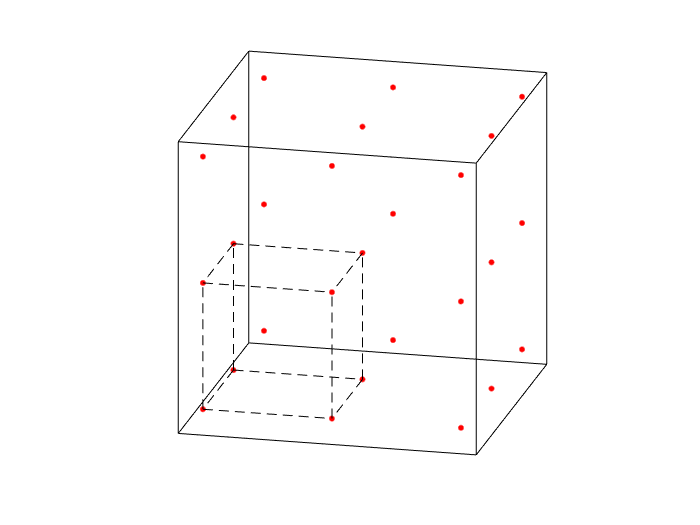}
    \captionsetup{justification=centering}
    \caption{Schematic of a cell employing $p = 2$ (third order accurate) and three Gauss-Legendre nodes per mesh element).}
    \label{fig: p2 cell subelement}
\end{subfigure}
\end{figure}

Figure \ref{fig: eta_eps} depicts the Gaussian projection function for different values of the parameter $\epsilon$. In Eq. \ref{eq: eta_eps} the parameter $\epsilon$ determines the width of the Gaussian projection. For very low $\epsilon_k$ we observe a numerical divergence.  For large $\epsilon$, the forces spread, leading to diffused blades and compromising the accuracy of the method. Different approaches have been investigated to determine its optimal value, such as $\epsilon = \max \left[ \frac{c}{4},\hspace{1mm} 4\Delta_{grid}, \hspace{1mm}\frac{c\cdot C_D}{2}\right]$, or more broadly, within a certain range $\epsilon \in [\Delta_{grid}, \hspace{1mm} 4\Delta_{grid}]$ \cite{Tossas_epsilon,Toldborg_epsilon}, where $\Delta_{grid} = (\Delta_x \Delta_y \Delta_z)^{\frac{1}{3}}$ represents the size of the cell used, $c$ refers to the airfoil chord and $C_{D}$ denotes the corresponding drag coefficient. 

In high-order $h/p$ solvers, there are $(p+1)^3$ (where $p$ is the polynomial order) Gauss-Legendre points contained within a mesh element. This enables defining the parameter $\epsilon_k$ within each element based on the polynomial order employed in each element. Figure \ref{fig: p2 cell subelement} shows a pictorial representation of subelements generated within a mesh element when a polynomial order $p=2$ (3rd order accuracy) is selected. Accordingly, the concept of $\Delta_{grid}$ can be redefined for high-order methods as follows:
\begin{equation}
    \Delta_{grid} = \frac{(\Delta_x \Delta_y \Delta_z)^{\frac{1}{3}}}{p+1}.
    \label{eq: Delta grid}
\end{equation}
%
Based on this definition, we can relate the Gaussian smoothing function to the effective grid spacing  $\Delta_{grid}$, by setting
\begin{equation}
\epsilon_k =k\times\Delta_{grid}.
    \label{eq: epsilon_k}
\end{equation}
Varying $k$ allows smearing the forcing between more high order nodes. For example $k=2$, distributed the forces in at least 2 high-order nodes. Finally, note that combining \ref{eq: Delta grid} and \ref{eq: epsilon_k}, we obtain:
\begin{equation}
\epsilon_k=k\times\Delta_{grid} = k\times\frac{(\Delta_x \Delta_y \Delta_z)^{\frac{1}{3}}}{p+1},
\end{equation}
where $k$ is a constant, usually within a certain range $k \in [1, \hspace{1mm} 4]$~\cite{Tossas_epsilon,Toldborg_epsilon} for low order methods. The last equation reflects that when the polynomial order is increased, the value of the projection parameter $\epsilon_k$ decreases (by definition), which is the pursued relationship between spatial discretization and the Gaussian smoothing.

In this work, we compare two methodologies AL1: averaged velocity and AL2: weighted sums. These two approaches result in different flow behaviors in the context of high-order $h/p$ solutions and are detailed hereafter.

\subsubsection{AL1: Cell averaged velocity}

In this strategy, we consider the average velocity in each mesh, which is computed via the high-order Gauss-Legendre nodes.

\begin{equation}
    \overline{\mathbf{q_t}} = \frac{1}{N}\sum_{i=1}^{N} \ \mathbf{q_t}_i; \quad N = (P+1)^3,
    \label{eq: averaged q}
\end{equation}
where the subscript $i$ denotes the node index within the element, $N$ represents the number of nodes, and $p$ is the order of the polynomial. 

At each time step, the positions of all actuator points are updated. Each actuator point is based on the averaged cell quantities of the conservative variables.
The average of conservative variables provides the flow velocity $\mathbf{v}$, which is used to calculate the components of the local velocity Eq.~\ref{eq: u_theta} and angle of attack Eq.~\ref{eq: angle of attack and local} at each actuator point and using Eq.~\ref{eq: Fx Fy Fz} to Eq.~\ref{eq: source_term_eqn}. To obtain the forces and corresponding source terms for all the nodes that are in the vicinity (defined by the Gaussian function $\eta_{d}$ and the parameter $\epsilon_k$) of the actuator point.

This procedure is equivalent to converting the solution that contains the actuator point into a first-order solution (as in finite volume methods). In principle, this procedure can avoid high-order frequencies when calculating local variables to compute the forces at the actuator point.

\subsubsection{AL2: Weighted sum forcing}

In this second approach, instead of averaging the state vector $\mathbf{q_t}$ to obtain the average velocity in the cell, we calculate an aerodynamic forcing in each high-order Gauss-Legendre node using the local flow variables. These forces are scaled using the Gaussian function $\eta_{d}$ and the parameter $\epsilon_k$. Subsequently, a weighted average (based on $\eta_{d}$) is used to calculate the general axial and tangential forces as:
\begin{equation}
    \overline{f}_j = \frac{\sum_{i=1}^{N} \eta_{ji}(d) \cdot f_i}{\sum_{j=1}^{N_a}\sum_{i=1}^{N}\eta_{ji}(d)},
    \label{eq: wieghted average}
\end{equation}
where $\eta_{ji}(d)$ represents the Gaussian projection from the actuator point $j$ to node $i$, $f_i$ denotes the contribution of node $i$ to the force, $\overline{f}_j$ represents the total force generated at the actuator point $j$, $N = (p+1)^3$ is the number of nodes in each element and $N_a$ the number of actuator points. The projection of forces onto the intrinsic axes of the problem and the addition to the Navier-Stokes equations as a source term are carried out in the same manner as in the AL1 method. 
Finally, let us note that if only one Gauss-Legendre point is available (polynomial $p=0$) the AL1 and AL2 are exactly equivalent, but as the polynomial increases, both formulations provide different forcing. Additionally, AL1 has a local formulation while AL2 can use information for multiple elements (the extension is controlled by the Gaussian shape and $\epsilon_k$).

\subsection{Computational costs of the AL1 and the AL2 implementations}

Before examining the results, we compare the computational cost of the two actuator line implementations AL1 and AL2. We run a simulation (after the initial transients) for $100$ time steps and compare the costs based on the polynomial order.
\begin{figure}[H]

  \centering
  \includegraphics[scale = 0.4]{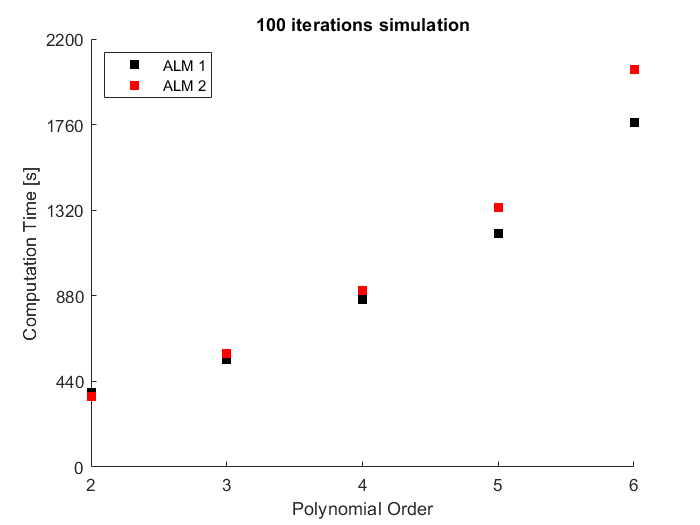}
\captionsetup{justification=centering}
\caption{Computational cost for the AL1 (averaging) and the AL2 (projection): 100 Iterations for polynomial orders ranging from 2 to 6.}
\label{fig: Time}

\end{figure}

Figure \ref{fig: Time}, shows the computational cost for the AL1 and AL2 implementations and for polynomial orders ranging from $p=2$ to 6. 
For low polynomials ($p<4$) both methods show comparable costs, but as the polynomial order increases, we observe
that AL2 becomes more costly than AL1. This overhead is attributed to the higher number of projections carried out in the latter (with respect to the former), which is proportional to the number of degrees of freedom $(p+1)^3$ included in each mesh element. This suggests that using the AL1 implementation may be more convenient at high polynomial orders if prioritizing the calculation time. However, as will be shown later, the AL1 implementation can introduce nonphysical oscillations for high polynomials, and therefore the AL2 implementation is recommended, despite its increased cost.

\section{Wind turbine definition and solver settings}\label{sec:definition}

We simulate the wind turbine experiment of the Norwegian University of Science and Technology (NTNU), which is a three-bladed wind turbine with a diameter $D$ of 0.894 m, which was tested in the wind tunnel of the Dept. Energy and Process Engineering, NTNU. The turbine blades were made up of NREL S826 airfoils \cite{Datos_experimentales}. 
For our analysis, we used the setup and data from the experiments, which provided information on the wake of a single wind turbine at various tip speed ratios $\lambda$. Specifically, we focus on the results at the optimal tip speed ratio $\lambda=\Omega D/2U_\infty = 6$, with $U_\infty=10$~m/s and $\Omega=134.228$~rad/s. The Reynolds number of the blade tip is $Re_c = \lambda U_{\infty} c_{tip}/\nu= 103600$, where $c_{tip}=0.025926$ m is the length of the tip chord and $\nu$ is the kinematic viscosity of the air. During the wind tunnel tests, the inflow velocity profile was uniform and the levels of turbulence intensity were low (below 0.3\%). 
The different airfoil cross sections along the blade, with their respective chord lengths and twist angles, can be found in \cite{Blind_Test_geometria}. The lift and drag coefficients are interpolated based on the angle of attack and Reynolds. XFOIL is used to generate the lift and drag polars for the NREL S826 airfoil profiles that define the wind turbine blades. We tabulate 31 blade sections with aerodynamic coefficients for a range of Reynolds numbers between 20k and 200k and interpolate on the fly to the required one.


The wind tunnel used for the experiments has the following dimensions:
\begin{center}
    Length (x): $L_x = 11.15 \hspace{1.5mm} m$;\quad Width (y): $L_y = 2.71 \hspace{1.5mm} m$;\quad Height (z): $L_z = 1.8 \hspace{1.5mm} m$
\end{center}



The fluid domain is discretized with a homogeneous mesh, resulting in $n_x = 128$ cells in the $x$ direction, while $n_y = n_z = 24$ cells in the $y$ and $z$ directions. With these data, we can calculate the values of $\Delta_{grid}$, following \ref{eq: Delta grid}, and $\epsilon_k =k\times\Delta_{grid}$, see \ref{eq: epsilon_k}, for varying $k$ and for each polynomial order ranging from $p=2$ to 5:

\begin{table}[H]
\hspace{-10mm}
\begin{tabular}{cccccccccc}
{p} &  {DOF}& \({\Delta_{grid} [m]}\) & \({\epsilon_{1.5}}\) & \({\epsilon_{2}}\) & \({\epsilon_{2.25}}\) & \({\epsilon_{2.5}}\) & \({\epsilon_{2.75}}\) & \({\epsilon_{3}}\) & \({\epsilon_{3.5}}\) \\
2    &  $1.99\times10^6$    & 0.03012        & 0.04518       & 0.06024     & 0.06777        & 0.07530       & 0.08283         & 0.09036     & 0.10542       \\
3   &   $4.72\times10^6$     & 0.02259        & 0.033885      & 0.04518     & 0.05083        & 0.05647       & 0.0621         & 0.06777     & 0.07906       \\
4    &   $9.22\times10^6$     & 0.01807        & 0.027105      & 0.03614     & 0.04066        & 0.04518       & 0.04969        & 0.05421     & 0.06325      \\
5    & $1.59\times10^6$      & 0.01506        & 0.02259       & 0.03012     & 0.03389        & 0.03765       & 0.04142        & 0.04518     & 0.05271
\end{tabular}
\captionsetup{justification=centering}
\caption{Summary of the meshes used and actuator line parameters, including the polynomial order $p$, the total number of degrees of freedom DOF and the associated values for the Gaussian smoothing $\Delta_{grid}$ and $\epsilon_k$ (which are based on the polynomial order $p$).}
\label{tab: epsilon}
\end{table}

\section{Results and discussion}\label{sec:results}


\subsection{Preliminaries: Simulations including the tower and the nacelle}\label{sez:preliminaries}
To mimic real turbines, the tower and nacelle should be included in the simulations. However, when including these bluff bodies in the flow we observe the generation of large turbulent structures that make difficult the study of the actuator lines.

\begin{figure}[H]

  \centering
  \includegraphics[trim={0.7cm 7cm 0.7cm 7cm}, clip,width=0.95\linewidth]{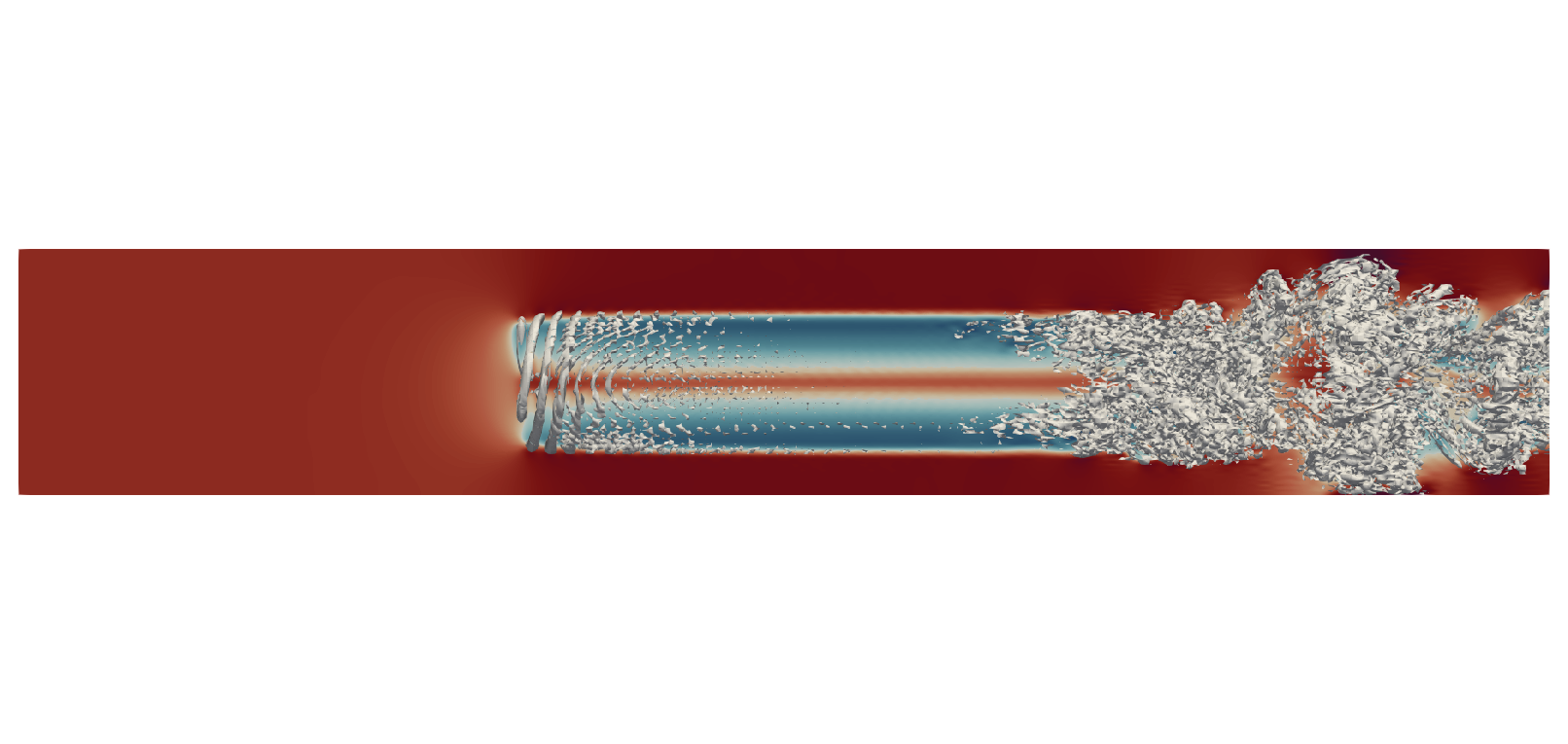}
 \vspace{-3mm}
 
 a) Actuator line without tower and nacelle.
 
  \includegraphics[trim={0.7cm 7cm 0.7cm 7cm}, clip,width=0.95\linewidth]{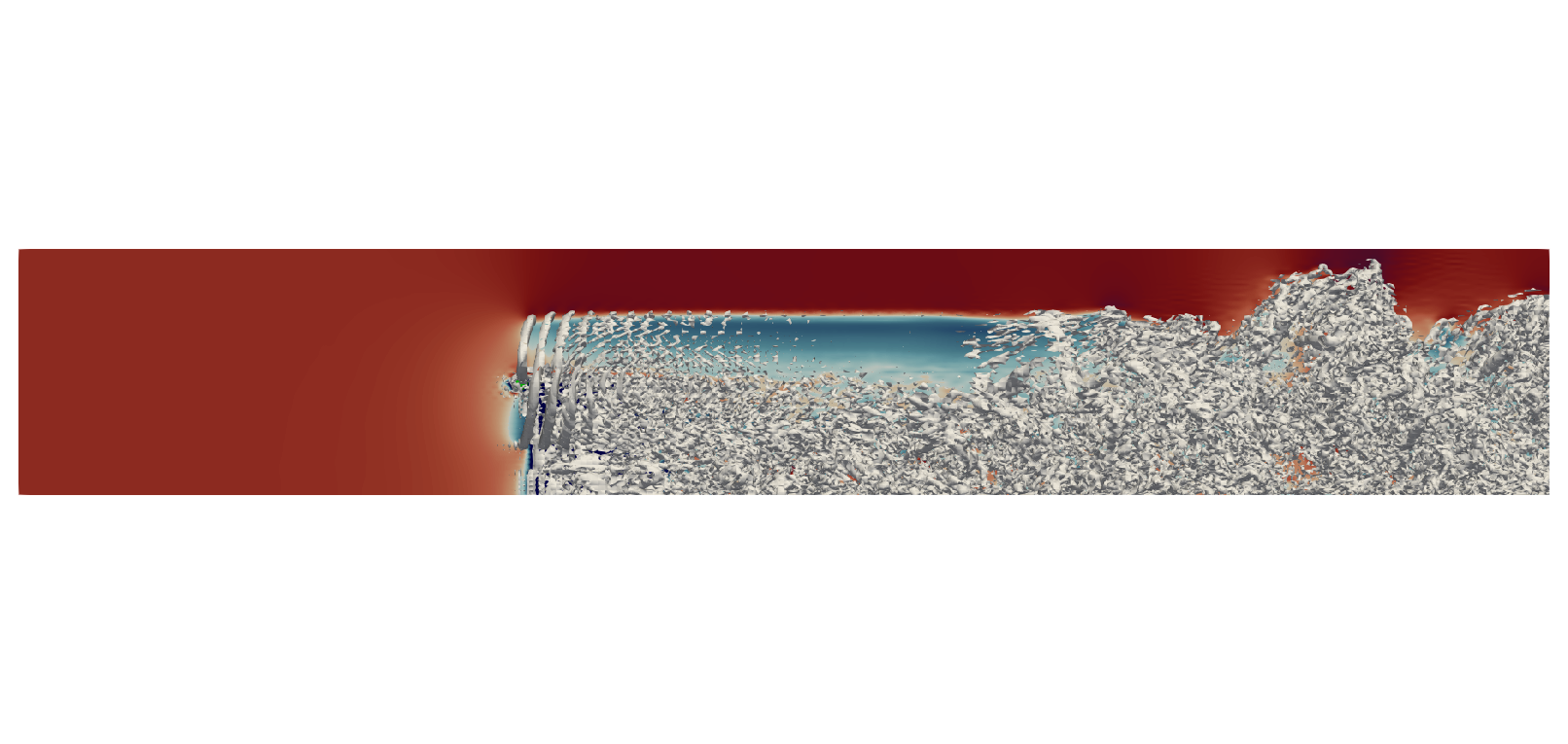}
\vspace{-3mm}

 b) Actuator line with tower and nacelle, which are modeled using immersed boundaries.
 
\captionsetup{justification=centering}
\caption{Instantaneous velocity field along the x/y-plane and  Q-criterion iso-surfaces:
a) Actuator line without tower and nacelle. b) Actuator line with tower and nacelle, modeled using immersed boundaries.}
\label{fig:u_qcrit_tower}

\end{figure}

Indeed, when incorporating the tower and nacelle into the simulations, we observe that the presence of the tower alters the flow field near the rotor, enhancing turbulence and mixing in the wake, influencing the turbine's performance and power output. Additionally, the nacelle creates aerodynamic interference, affecting airflow over the rotor blades and altering the aerodynamic forces experienced by the turbine. These effects are clearly visible through the Q-criterion in figures \ref{fig:u_qcrit_tower}, where we compare simulations with/without nacelle and tower. Here, the nacelle and tower are modeled using immersed boundaries \cite{KOU2022110721,HORSES_3D}, which allows explicitly including the three-dimensional shape of the objects. Note that the tower and nacelle are not represented as a flow resistance object via discrete source terms (as in the actuator line method). 

This preliminary ALs simulations (with and without immersed boundaries) are performed using a h-mesh with element size half size in each direction as the one described before ($n_x=256, n_y=n_z=48$) and polynomial order 2, such that the total number of degrees of freedom is the same as the cases with polynomial order 5 in the mesh originally described.

\begin{figure}[H]
    \centering
    \includegraphics[trim={0cm 0cm 0cm 0.92cm}, clip,width=0.48\linewidth]{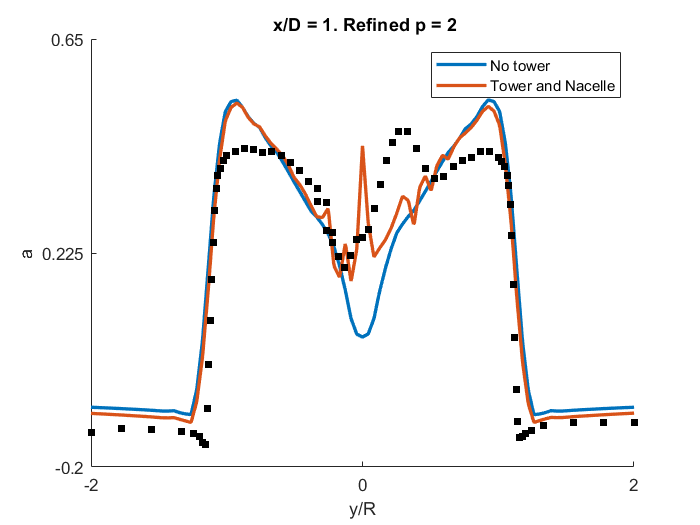}
    \label{fig:tower a vs y 1}
    \includegraphics[trim={0cm 0cm 0cm 0.92cm}, clip,width=0.48\linewidth]{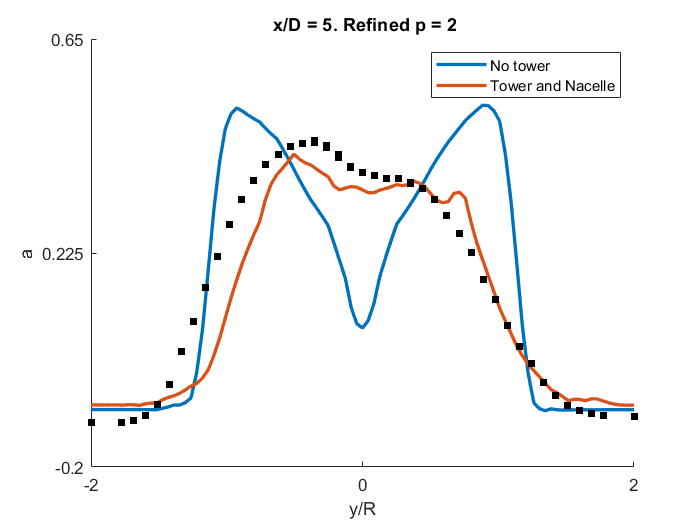}
    \captionsetup{justification=centering}
    \caption{Velocity decay, $a$, with a $p = 2$ h-mesh with and without tower and nacelle implementation. Left $x/D=1$, right $x/D=3$. \\'---' AL1, '$\blacksquare$' Experimental data.}
    \label{fig:tower a vs y 3}
\end{figure}

Incorporating these geometries in the simulations allows for more accurate predictions of power outputs, load distributions, and wake deficits. Figure \ref{fig:tower a vs y 3} shows the wake deficit $a = 1 - \frac{U}{U_\infty}$ at two planes behind the rotor ($x/D=$1 and 5). We observe that including the tower and nacelle allows to better capture the experimental results and particularly far from the rotor at x/D=5 where very good agreement is obtained. When comparing the wakes with and without the tower and the nacelle, observe that the symmetry of the wake is lost when including these additional geometries. We conclude that including the tower and nacelle includes leads to a more rich flow and complex wake, but can hide small differences when modelling rotating blades using ALs. For this reason, we have preferred not to include the tower and the nacelle in the following sections. By doing so, small differences between the actuator line implementations will be easily identified.

\subsection{Power and thrust coefficients}\label{sec:Cp_Ct}

The Blade Element Momentum Theory \cite{BEMT} provides the definitions of the power coefficient ($C_P$) and the thrust coefficient ($C_T$):
\begin{equation}
    C_P = \frac{Q\Omega}{\frac{1}{2}\rho V^3 \pi R^2}; \quad C_T = \frac{T}{\frac{1}{2}\rho V^2 \pi R^2}.
    \label{eq: cp ct}
\end{equation}
where $Q$ represents the total torque of the wind turbine and $T$ represents the thrust, both non-dimensional with respect to the maximum available power and thrust, based on the rotor frontal area. The experimental values for the wind tunnel are $C_P = 0.45$ and $C_T = 0.9$. Tables \ref{tab: CP CT} presents the results for the simulations performed (without tower and nacelle) for varying Gaussian shapes $\epsilon_k$ and polynomial orders ranging from $p=2$ to 5, and for the formulations AL1 (averaging) and AL2 (weighted sums). The tables include the time-averaged coefficient values. 
The averaging process is started once the coefficient values reach a statistically steady state. It can be observed that both coefficients show an increasing trend when the projection parameter $\epsilon_k$ increases and for a fixed polynomial order.

\begin{table}[H]
\hspace{-6mm}
\begin{minipage}{.55\textwidth}
  \centering
  \begin{tabular}{c|cccc}
  \multicolumn{5}{l}{\textbf{Actuator line AL1} }\\
    \hline
  \multicolumn{5}{c}{\textbf{p=2} }\\           \hline
    $\epsilon_k$ & $\epsilon_2$ & $\epsilon_{2.25}$ & $\epsilon_{2.5}$ & $\epsilon_{2.75}$ \\
    $C_P$                    & 0.589478                   & 0.604494                      & 0.62109                      & 0.637955                      \\
    $C_T$                    & 0.899552                   & 0.90732                       & 0.916623                     & 0.925953                      \\ \hline
   \multicolumn{5}{c}{\textbf{p=3} }\\          \hline
    $C_P$                    & -                          & -                             & 0.58497                      & 0.595791                      \\
    $C_T$                    & -                          & -                             & 0.899046                     & 0.903251                      \\ \hline
   \multicolumn{5}{c}{\textbf{p=4} }\\    \hline
    $C_P$                    & -                          & -                             & -                            & 0.574034                      \\
    $C_T$                    & -                          & -                             & -                            &  0.891353              \\ \hline
  \multicolumn{5}{c}{\textbf{p=5} }\\     \hline
    $C_P$                    & -                          & -                             & -                            & 0.564624                      \\
    $C_T$                    & -                          & -                             & -                            & 0.887237 
  \end{tabular}
  \label{tab: CP CT AL1}
\end{minipage}%
\begin{minipage}{.475\textwidth}
  \begin{tabular}{cccc}
    \multicolumn{4}{l}{\textbf{Actuator line AL2} }\\
    \hline
  \multicolumn{4}{c}{\textbf{p=2} }\\     \hline
     $\epsilon_2$ & $\epsilon_{2.25}$ & $\epsilon_{2.5}$ & $\epsilon_{2.75}$ \\
     0.628962                   & 0.64718                      & 0.666427                      & 0.684074                      \\
    0.904623                   & 0.916546                       & 0.929052                     & 0.939782                      \\ \hline
  \multicolumn{4}{c}{\textbf{p=3} }\\     \hline
     -                          & -                             & 0.621007                      & 0.63465                      \\
     -                          & -                             & 0.899047                     & 0.908509                      \\ \hline
  \multicolumn{4}{c}{\textbf{p=4} }\\     \hline
     -                          & -                             & -                            & 0.606609                      \\
     -                          & -                             & -                            & 0.888485              \\ \hline
  \multicolumn{4}{c}{\textbf{p=5} }\\     \hline
     -                          & -                             & -                            & 0.594947                      \\
    -                          & -                             & -                            & 0.875988                  
  \end{tabular}
  \label{tab: CP CT AL2}
\end{minipage}
\captionsetup{justification=centering}
\caption{AL1 averaging method and AL2 weighted sums method: $C_P$ and $C_T$ based on the polynomial order and $\epsilon_k$ for $k\in[2,3]$.}
\label{tab: CP CT}
\end{table}

$C_P$ and$C_T$ compare well with the experimental values (given the absence of the tower and the nacelle).
Increasing $\epsilon_k$ increases the power and thrust coefficients, while increasing the polynomial (for fixed $\epsilon_k$) decreases the coefficients. Both AL1 and AL2 results converge as we increase the polynomial order (see fixed $\epsilon_{2.75}$), while no clear convergence is observed when modifying $\epsilon_k$ (for fixed $P=2$). This suggests that $\epsilon_k$ changes the model of the actuator line and is thus related to a modelling error and not a discretization error (which is the case for the polynomial order).
Additionally, we note that the AL2 implementation provides higher values (w.r.t the AL1 implementation) for $C_P$ and $C_T$ and for low polynomials and all $\epsilon_k$. However, for high polynomials (see $P=5$), the AL2 provides similar values to Al1. The last observation can be explained when analysing the the force distributions along the actuator lines, see the next section.

\subsection{Force distributions along the blade}

In this section we compare the 
the time-averaged axial and tangential forces along one blade for changes in polynomial order  and varying Gaussian smoothing $\epsilon_k$ with $k \in [1, \hspace{1mm} 3]$.

\begin{figure}[H]

\begin{subfigure}{0.5\textwidth}
  \centering
  \includegraphics[scale = 0.4]{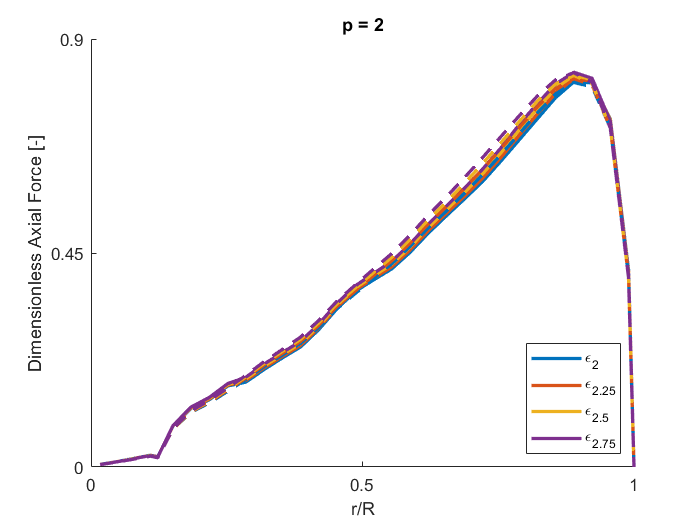}
  
  a)
\end{subfigure}
\hfill
\begin{subfigure}{0.5\textwidth}
  \centering
  \includegraphics[scale=0.4]{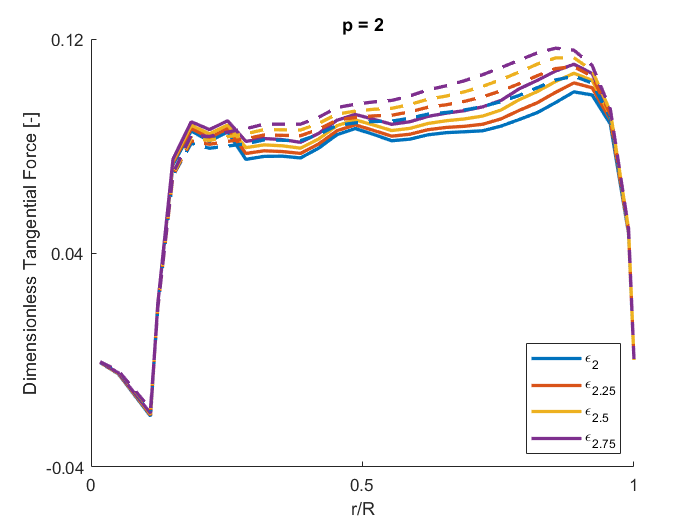}
  
  b)
\end{subfigure}

\begin{subfigure}{0.5\textwidth}
  \centering
  \includegraphics[scale = 0.4]{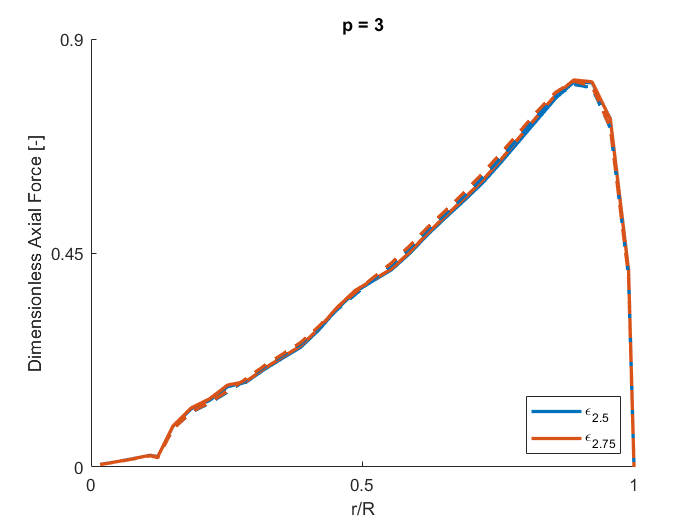}
  
  c)
\end{subfigure}
\hfill
\begin{subfigure}{0.5\textwidth}
  \centering
  \includegraphics[scale = 0.4]{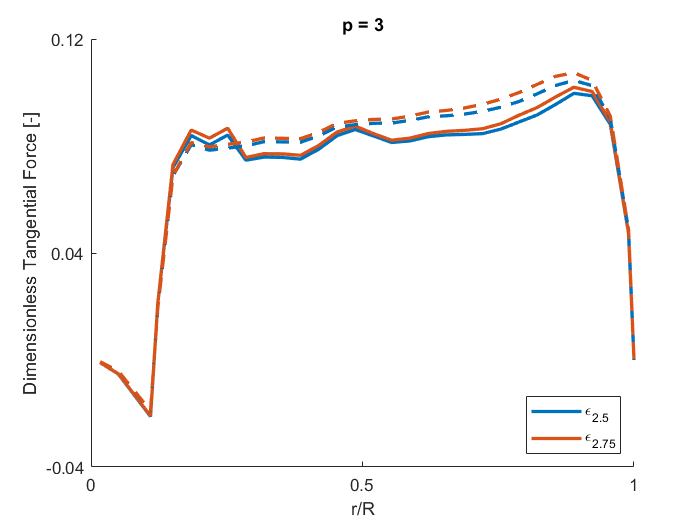}
  
  d)
\end{subfigure}

\begin{subfigure}{0.5\textwidth}
  \centering
  \includegraphics[scale = 0.4]{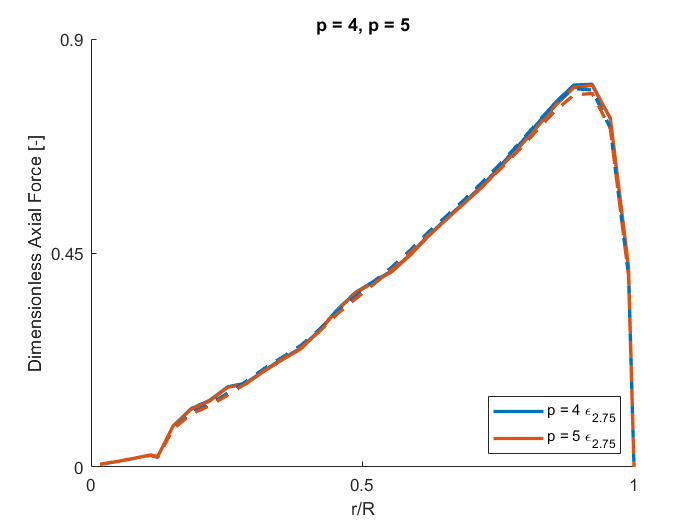}
  
  e)
\end{subfigure}
\hfill
\begin{subfigure}{0.5\textwidth}
  \centering
  \includegraphics[scale = 0.4]{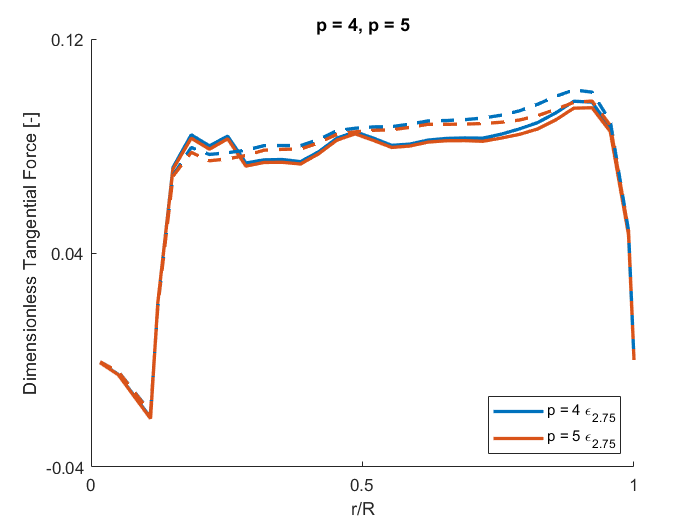}
  
  f)
\end{subfigure}
\captionsetup{justification=centering}
\caption{Axial and tangential non-dimensional forces along the blade based on polynomial order. \\  '---' AL1 '- -' AL2.}
\label{fig: blade forces p2}
\end{figure}

Figure \ref{fig: blade forces p2} depicts the time-averaged of the dimensionless axial and tangential forces along one blade. We include results for AL1 (continuous line) and AL2 (dashed line) and compare polynomial orders $P=2$ to 5, together with changes in Gaussian smoothing.  A sharp change in the tangential force can be observed around $r/R \approx 0.1$ in Figures \ref{fig: blade forces p2} b), d) and f), which is attributed to the abrupt change in the sectional profiles in the model definition \cite{Blind_Test_geometria} (transition from the circular profile at the root to the NREL-S826 airfoil for the rest of the blade). Beyond this point, we encounter a homogeneous distribution for the tangential force. 
Figures \ref{fig: blade forces p2} a), c), and e) show that in the axial force there is a monotonic increase as we approach the blade tip, resulting from the increase in local velocity due to rotation (and consequently, aerodynamic forces) 
Both axial and tangential forces decrease to zero at $r/R \approx 1$ due to the tip correction \cite{Tip_correction} introduced in both AL models. Similar trends for blade forces using ALs can be observed in previous works; see, for example, \cite{Datos_numericos}.

Increasing the polynomial order does not significantly affect the force distribution along the blades, while varying the Gaussian smoothing does provide slightly different distributions.
  If we compare AL1 and AL2, we see that AL2 yields a smoother and larger force distribution over the blade, compared to the AL1 method. The higher forces in AL2, for low polynomials, explain the larger power coefficients $C_P$ and the thrust coefficient $C_T$ observed in section \ref{sec:Cp_Ct}. 

\begin{figure}[H]

\begin{subfigure}{0.5\textwidth}
  \centering
  \includegraphics[scale = 0.4]{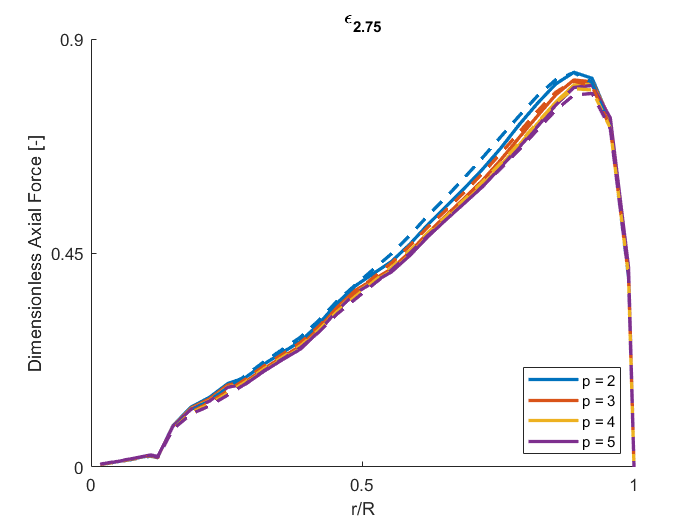}

  c)
\end{subfigure}
\hfill
\begin{subfigure}{0.5\textwidth}
  \centering
  \includegraphics[scale=0.4]{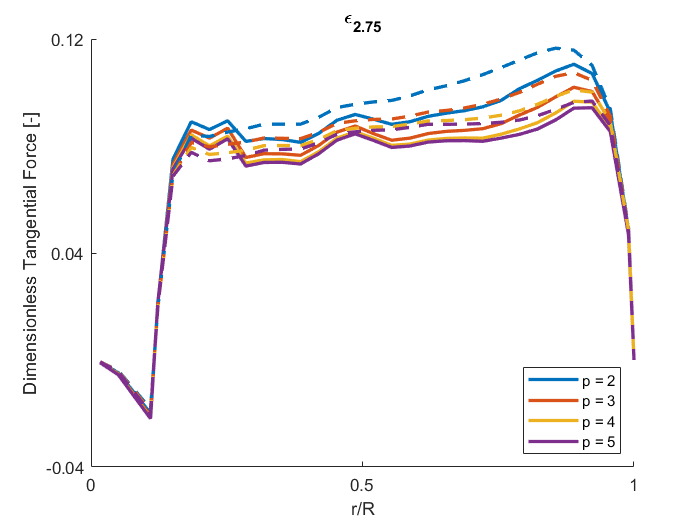}

  b)
\end{subfigure}
\captionsetup{justification=centering}
\caption{ Axial and tangential forces along the blade, $\epsilon_{2.75}$. \\  '---' AL1 '- -' AL2.}
\label{fig: forces e275}

\end{figure} \vspace{-2mm}

\begin{figure}[H]

\begin{subfigure}{0.5\textwidth}
  \centering
  \includegraphics[scale = 0.4]{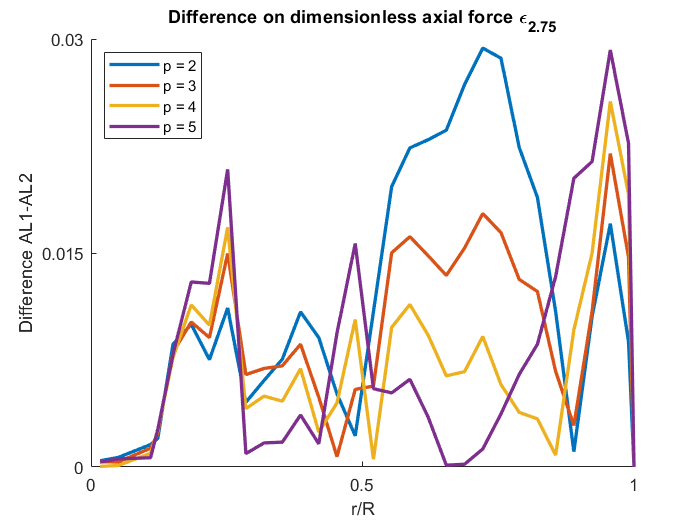}

  a)
\end{subfigure}
\hfill
\begin{subfigure}{0.5\textwidth}
  \centering
  \includegraphics[scale=0.4]{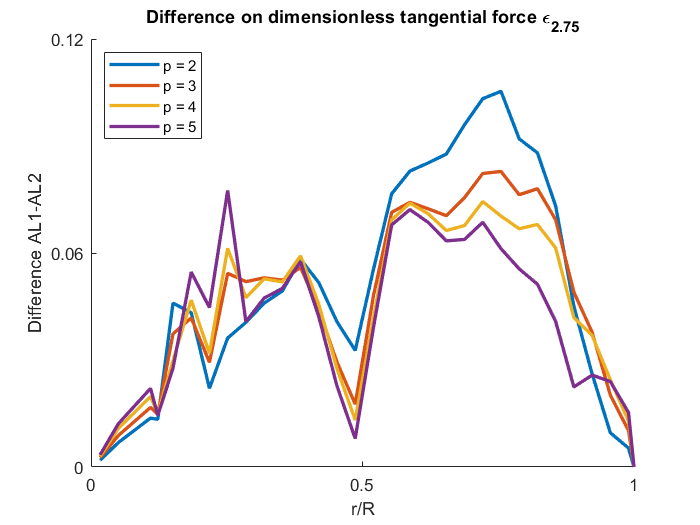}

  b)
\end{subfigure}
\captionsetup{justification=centering}
\caption{Relative difference in forces between actuator line methods (AL1-AL2) along the radius, and for  polynomial orders $P=2$ to 5:  a) axial and b) tangential forces.}
\label{fig: diff forces p}
\end{figure}

Figure \ref{fig: diff forces p} shows the differences in the forces between the AL1 and AL2 methods (and relative to the maximum force in the AL2) for axial and tangential forces. As the polynomial order increases, the two AL methods become more similar but remain different (the difference never reduces to zero).
The error in the tangential force is greater because the AL2 method provides a more homogeneous and smoother distribution (see Figure \ref{fig: forces e275}), while oscillations are present in the AL1 distributions. 

\begin{figure}[H]

\begin{subfigure}{0.5\textwidth}
  \centering
  \includegraphics[scale = 0.4]{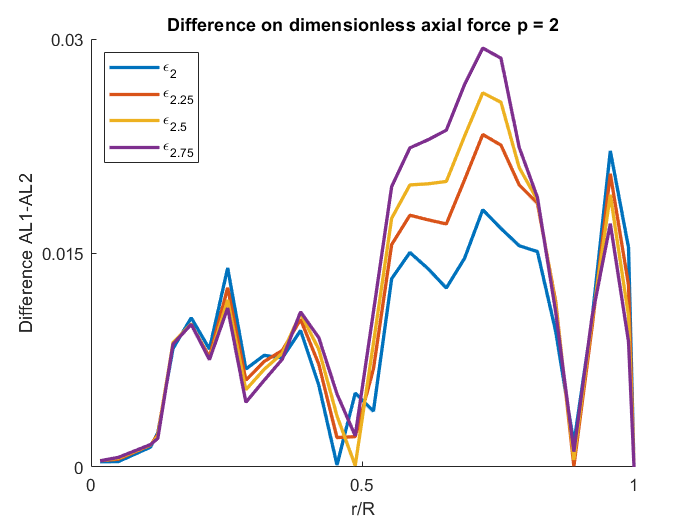}

  a)
\end{subfigure}
\hfill
\begin{subfigure}{0.5\textwidth}
  \centering
  \includegraphics[scale=0.4]{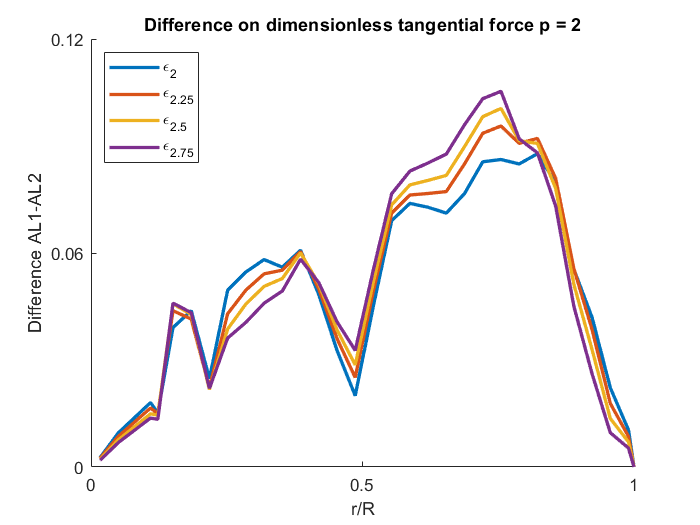}

  b)
\end{subfigure}
\captionsetup{justification=centering}
\caption{Relative difference in forces between methods as a function of $\epsilon_k$.}
\label{fig: diff forces eps}
\end{figure}

To isolate the effects of varying the projection parameter $\epsilon_k$, the polynomial order is set to $p = 2$, and $\epsilon_k$ is varied over a range of values $k\in[2,3]$ in
Figure \ref{fig: diff forces eps}. We again show the difference AL1-AL2.
We observe a smaller influence when varying the Gaussian smoothing, than when varying the polynomial order; see Figure \ref{fig: diff forces p}). However, the difference never reaches zero, showing that the two methods are modelling the forces differently, regardless of the Gaussian smoothing and the polynomial order.

%

\subsection{Velocity decay along the blade}

\begin{figure}[H]
   \begin{subfigure}{1\textwidth}
    \centering
    \includegraphics[width=0.3\textwidth]{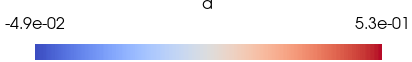}
     \end{subfigure}
    \begin{subfigure}{0.45\textwidth}
        \centering
        \includegraphics[width=\textwidth]{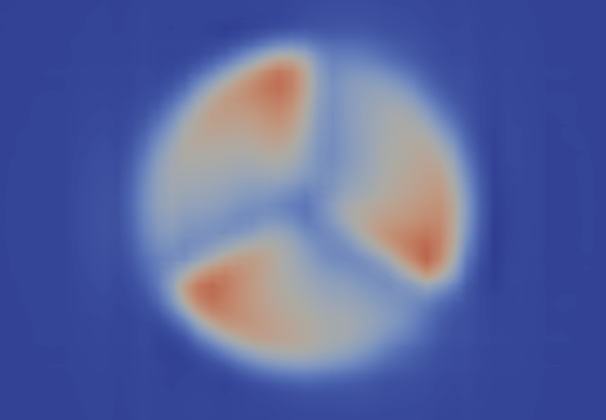}
        a)
    \end{subfigure}\hfill
    \begin{subfigure}{0.45\textwidth}
        \centering
        \includegraphics[width=\textwidth]{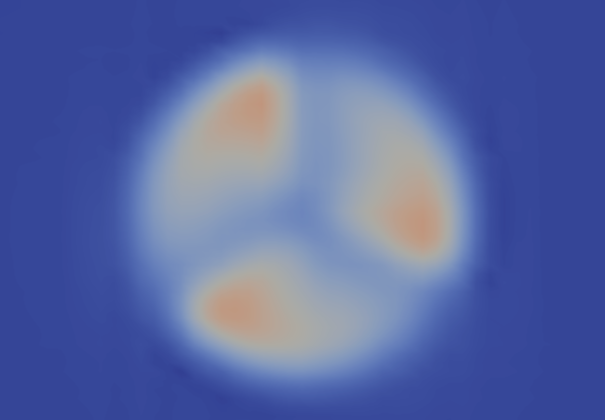}
        b)
    \end{subfigure}

\captionsetup{justification=centering}
    \caption{Velocity decay $a =1 - \frac{U}{U_\infty}$, at the rotor for polynomial order  $p=2$: a) $\epsilon_2$,
    b) $\epsilon_{2.75}$. Only the AL1 formulation is shown since for $p=2$ the distributions are almost identical for the AL2.}
    \label{fig: rotor eps}
\end{figure}

The forces presented in the previous section are introduced into the flow to create the interaction between the rotor and the air. 
To characterize the velocity decay at the rotor plane ($y-z$ axis), we define the velocity deficit $a =1 - \frac{U}{U_\infty}$, which is often referred to as the axial induction factor. Figure \ref{fig: rotor eps} shows the velocity deficit in the rotor plane for polynomial order $p=2$ and two values of Gaussian smoothing $\epsilon_k$, with $k=2$ and 2.75. We observe a smeared velocity distribution at the blade location when increasing the Gaussian smoothing, which is expected as the forces are spread over more high-order nodes. For low polynomials, this effect is mild, since there is a small number of high order nodes per mesh element (e.g., for $p=2$ only 3 Gauss nodes). 

\begin{figure}[H]
    \centering

    \includegraphics[scale=0.5]{figures/Results/a_vs_r/leyenda_rotor_a.png}\vspace{1mm}
    
    \begin{minipage}{0.45\textwidth}
        \centering
        \includegraphics[width=\textwidth]{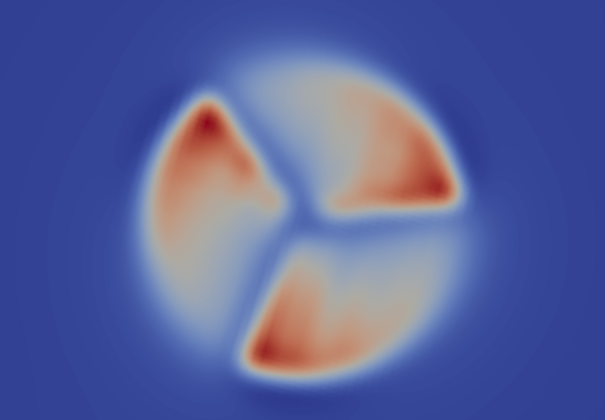}
        a)
    \end{minipage}\hfill
    \begin{minipage}{0.45\textwidth}
        \centering
        \includegraphics[width=\textwidth]{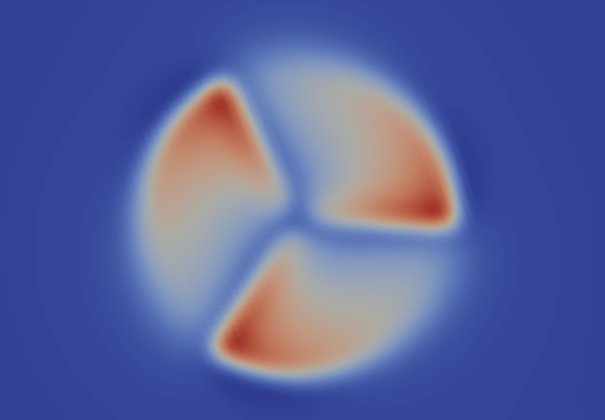}
        b)
\end{minipage}
\captionsetup{justification=centering}
\caption{Dimensionless velocity decay contours, $a =1 - \frac{U}{U_\infty}$, at the rotor for polynomial order $p = 5$ and $\epsilon_2.75$: a) AL1 method, 
    b)AL2 method.}
    \label{fig: rotor p5}
\end{figure}

Now we consider $p=5$ where differences can be seen for AL1 and AL2. Figure \ref{fig: rotor p5} shows the velocity decay at the rotor plane for $p=5$ and $\epsilon_2.75$. Consistently with the distribution of the blade forcing, the velocity deficit are smoother and more homogeneous for the AL2 implementation (than for the AL1), and should lead to a locally more uniform wake (see next sections). 

\begin{figure}[H]

    \begin{subfigure}{0.5\textwidth}
        \centering
        \includegraphics[scale = 0.35]{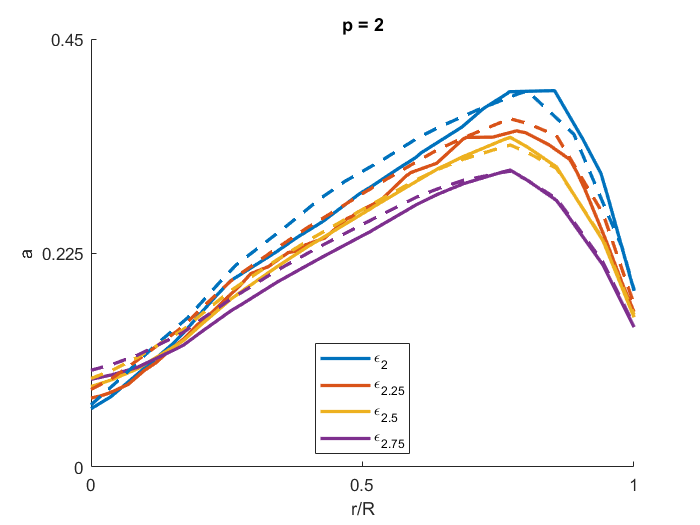}

        a)
    \end{subfigure}
        \begin{subfigure}{0.5\textwidth}
  \centering
  \includegraphics[scale = 0.35]{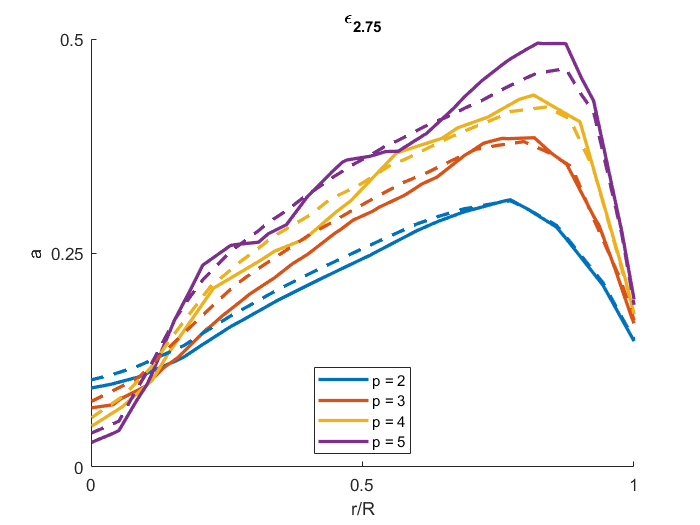}
  
  b)
   \end{subfigure}

\captionsetup{justification=centering}
\caption{Velocity decay along one blade for: a) varying $\epsilon_k$ and fixed polynomial $P=2$ and b) fixed $\epsilon_k$ and polynomial orders p02 to 5.  \\  '---' AL1'- -' AL2.}
\label{fig: a vs r}
\end{figure}

To quantify the above observations, Figure \ref{fig: a vs r} shows the axial wake deficit in the rotor and along the blade for a) changes in Gaussian smoothing (fixed $p=2$) and b) changes in the polynomial order (fixed $\epsilon_{2.75}$), and for the AL1 and AL2 implementations. We see that the wake deficit correlates well with the axial forces on the blades analyzed in Figures \ref{fig: blade forces p2} and \ref{fig: forces e275}. That the polynomial order induces larger changes in the velocity deficit than the Gaussian smoothing. Additionally, we confirm that the AL1 and AL2 implementations provide very similar results for low polynomial orders ($p=2$ and 3). For high polynomial orders ($p=4$ and 5) the AL2 formulation provides smoother velocity deficit distributions than the AL1 formulation. This effect will be investigated further in the following sections. 


\subsection{Time averaged wake velocity}

\begin{figure}[H]
    \centering
    \includegraphics[scale=0.40]{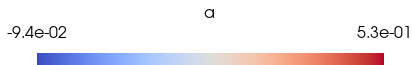}
    \vspace{0mm}
    
    \includegraphics[scale=0.402]{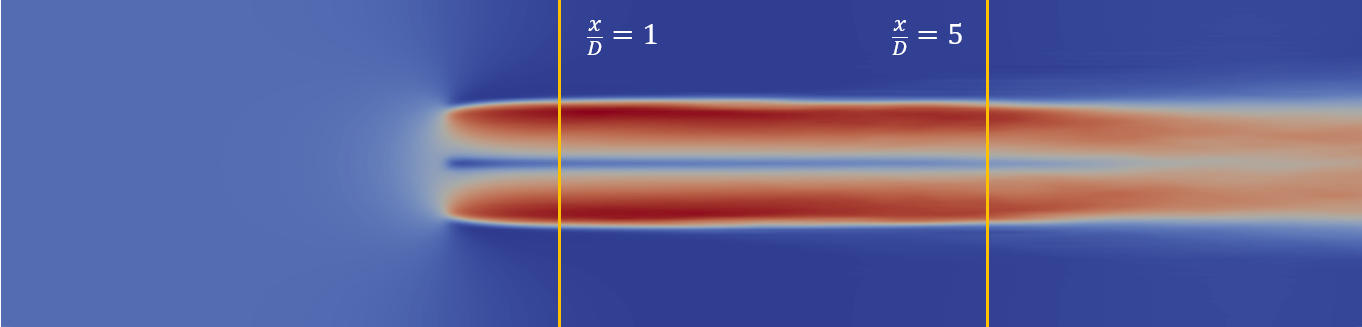}
    
    \vspace{-14.5mm}
    \begin{figure}[H]
  \hspace{18mm}
  \begin{tikzpicture}
    \draw[->,-{Latex[length=2mm]}] (0,0) -- (0.75,0) node[above] {$x$};
    \draw[->,-{Latex[length=2mm]}] (0,0) -- (0,0.75) node[right] {$y$};
  \end{tikzpicture}
\end{figure} \vspace{-8mm}

\caption{Planes in which the velocity decay, $a$, is represented.}
    \label{fig: wake segments x/D}
\end{figure}

\begin{figure}[H]

\vspace{-4mm}
\begin{equation}    \uline{\textbf{\normalsize{x\hspace{1mm}/\hspace{1mm}D = 1}}} \notag
\end{equation}
\vspace{-5mm}

\begin{subfigure}{0.5\textwidth}
  \centering
  \includegraphics[scale = 0.4]{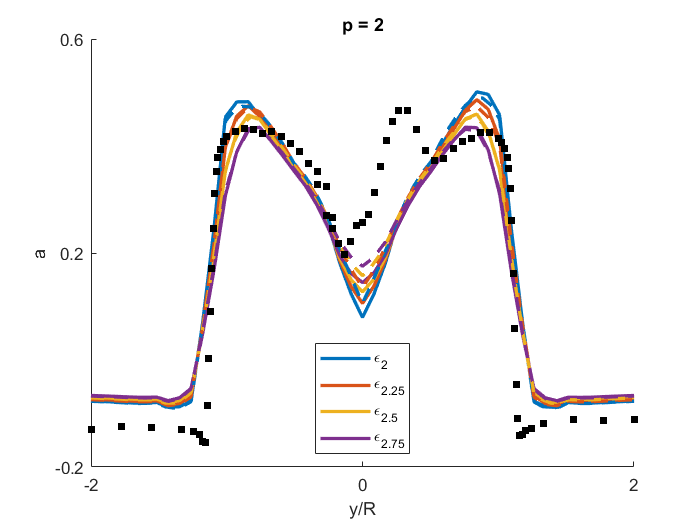}

    a)
  
\end{subfigure}
\hfill
\begin{subfigure}{0.5\textwidth}
  \centering
  \includegraphics[scale=0.4]{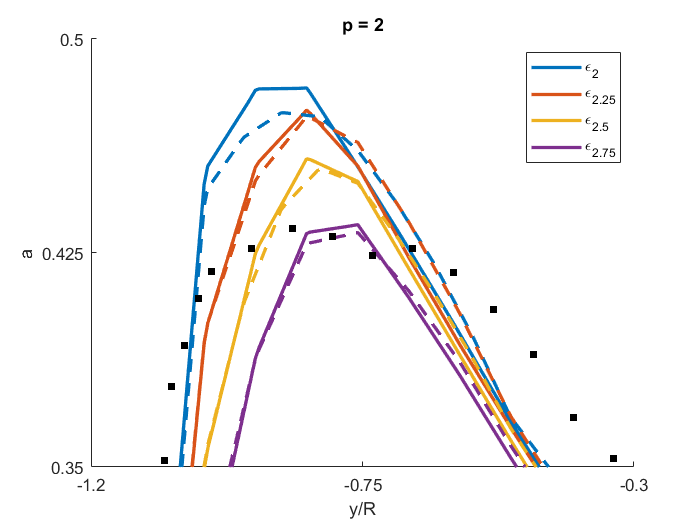}

    b)
  
\end{subfigure}

\begin{subfigure}{0.5\textwidth}
  \centering
  \includegraphics[scale = 0.4]{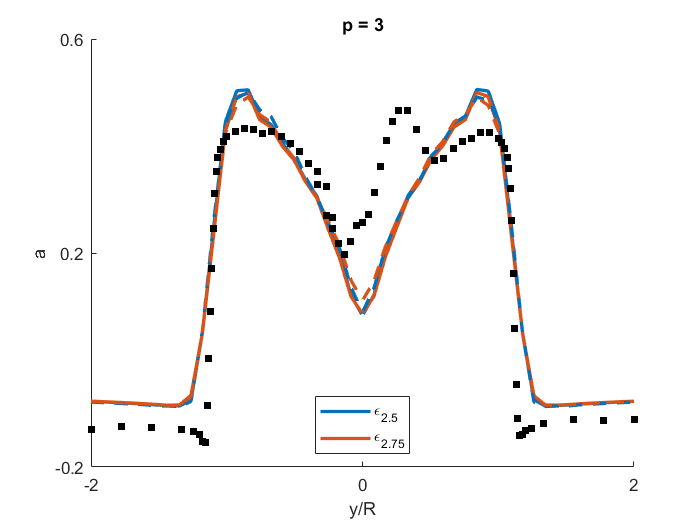}

    c)
  
\end{subfigure}
\hfill
\begin{subfigure}{0.5\textwidth}
  \centering
  \includegraphics[scale=0.4]{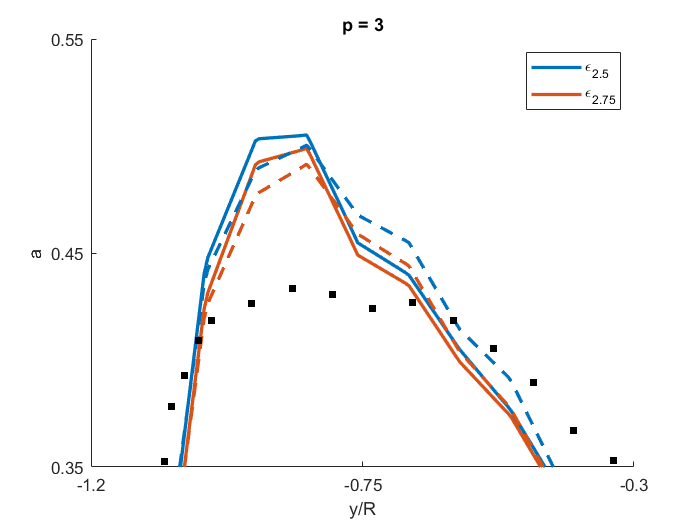}

    d)
  
\end{subfigure}

\begin{subfigure}{0.5\textwidth}
  \centering
  \includegraphics[scale = 0.4]{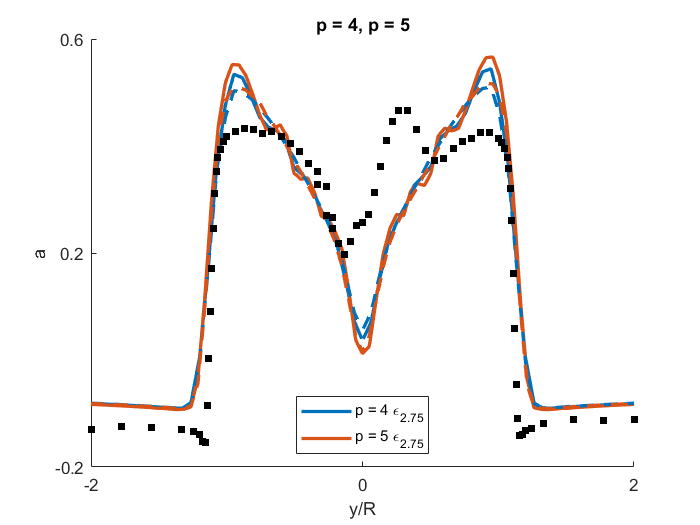}

    e)
  
\end{subfigure}
\hfill
\begin{subfigure}{0.5\textwidth}
  \centering
  \includegraphics[scale=0.4]{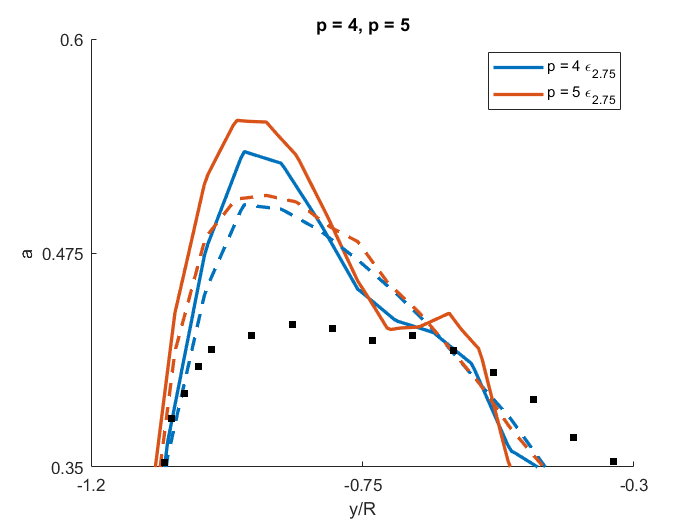}

    f)
  
\end{subfigure}
\captionsetup{justification=centering}
\caption{Velocity decay, $a$, for different polynomial order based on $\epsilon_k$, $ x/D = 1$. \\  '---' AL1 '- -' AL2 '$\blacksquare$' Experimental data. Right figures show zoomed detail of the wake peak.}
\label{fig: a vs y 1}

\end{figure}

In this section, we show the averaged velocity deficit $a$ in two streamwise planes in the wake of the turbine, aft of the rotor at distances $x/D=1$ and 5, as shown in Figure \ref{fig: wake segments x/D}. 
Figures \ref{fig: a vs y 1}.a-f) compare the time-averaged velocity for AL1 and AL2, including polynomial orders 2 to 5 and varying Gaussian smoothing.

\vspace{-6mm}

Figures \ref{fig: a vs y 1}.a) to d) show that for low polynomial orders ($p\leq3$), we observe a very similar behavior of the AL1 and AL2 methods. The effect of changing the smoothing parameter is similar in both methods. When increasing the polynomial to $p>3$, Figure \ref{fig: a vs y 1}.c) and f), we observe differences between the formulations AL1 and AL2. In particular for the AL1 oscillations appear in the averaged velocity.
When the Gaussian smoothing $\epsilon_k$ is reduced, the curve becomes sharper near the rotor edge (Figure \ref{fig: eta_eps}), illustrating that the flow has a greater deceleration when reducing $\epsilon_k$.  At the edge of the rotor $y/D/2 \approx 0.8$, there is an excess of the force introduced, compared to the experiments.  A sharper velocity has also been observed in the work of M.Zormpa et al. \cite{Oxford_sim} and is attributed to the decreased production of turbulent kinetic energy production near the rotor (reducing mixing), see next section.  Discrepancies near the hub are related to the missing nacelle in the simulations (see Figure \ref{fig:tower a vs y 3} and related comments).

\begin{figure}[H]
\begin{equation}
    \uline{\textbf{\normalsize{x\hspace{1mm}/\hspace{1mm}D = 5}}} \notag
\end{equation} \vspace{-5mm}

\begin{subfigure}{0.5\textwidth}

  \centering
  \includegraphics[scale = 0.4]{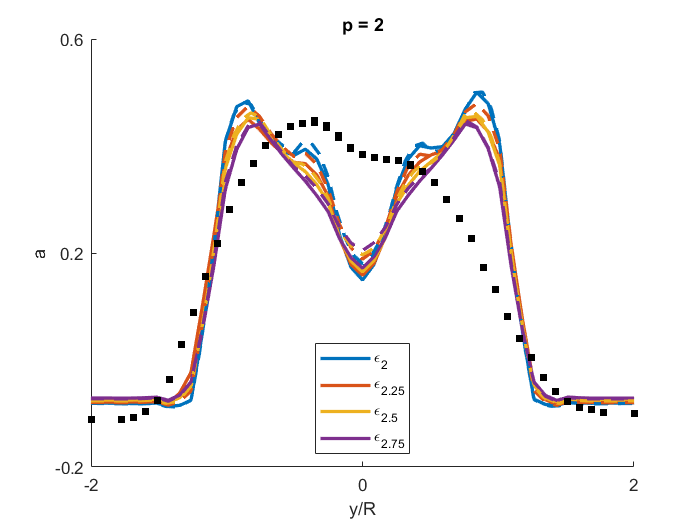}

    a)
  
\end{subfigure}
\hfill
\begin{subfigure}{0.5\textwidth}
  \centering
  \includegraphics[scale=0.4]{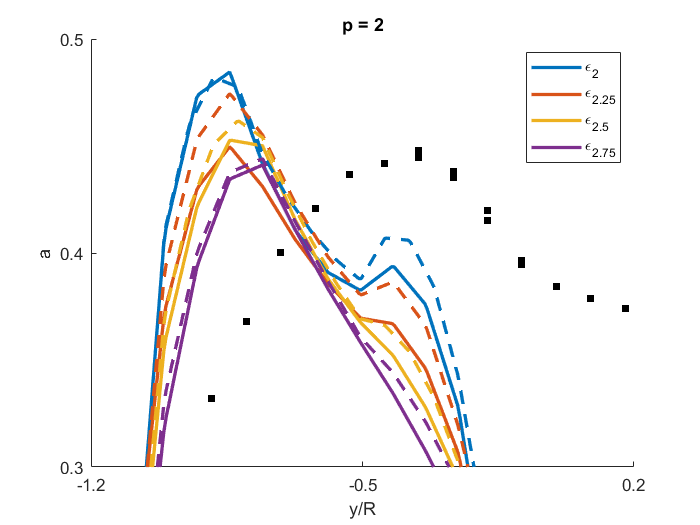}

    b)
  
\end{subfigure}

\begin{subfigure}{0.5\textwidth}
  \centering
  \includegraphics[scale = 0.4]{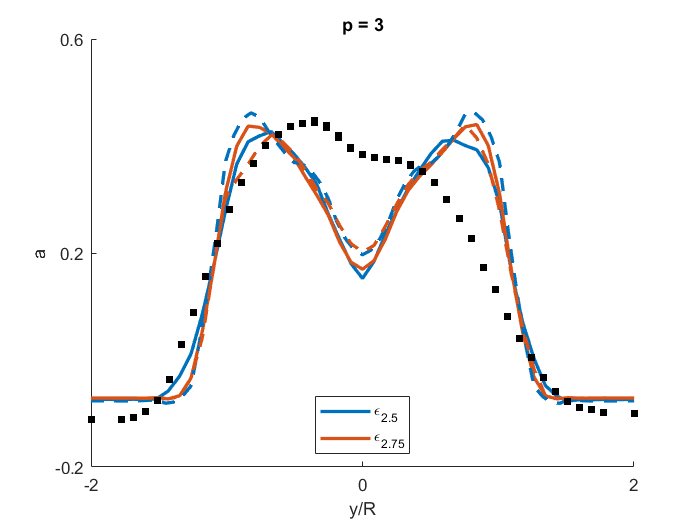}

    c)
  
\end{subfigure}
\hfill
\begin{subfigure}{0.5\textwidth}
  \centering
  \includegraphics[scale=0.4]{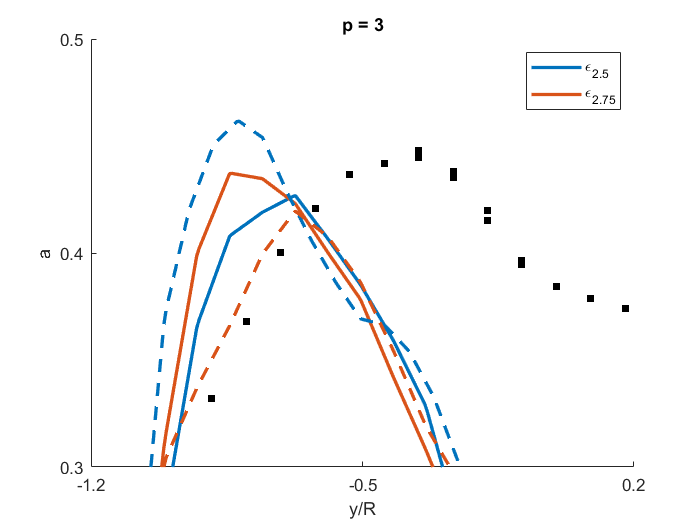}

    d)
  
\end{subfigure}

\begin{subfigure}{0.5\textwidth}
  \centering
  \includegraphics[scale = 0.4]{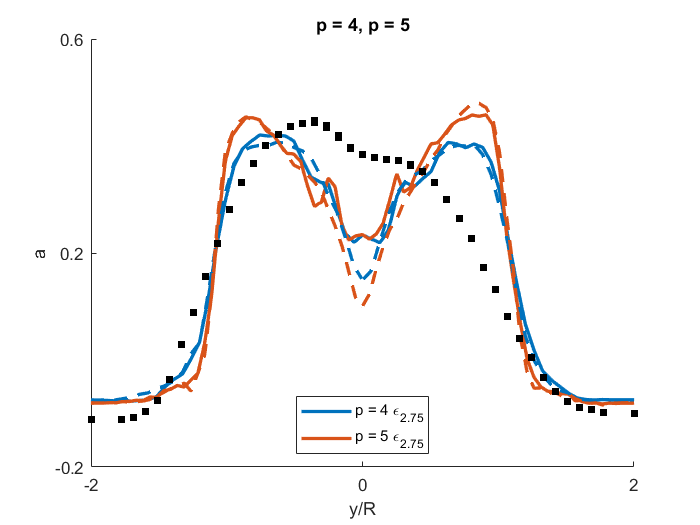}

    e)
  
\end{subfigure}
\hfill
\begin{subfigure}{0.5\textwidth}
  \centering
  \includegraphics[scale=0.4]{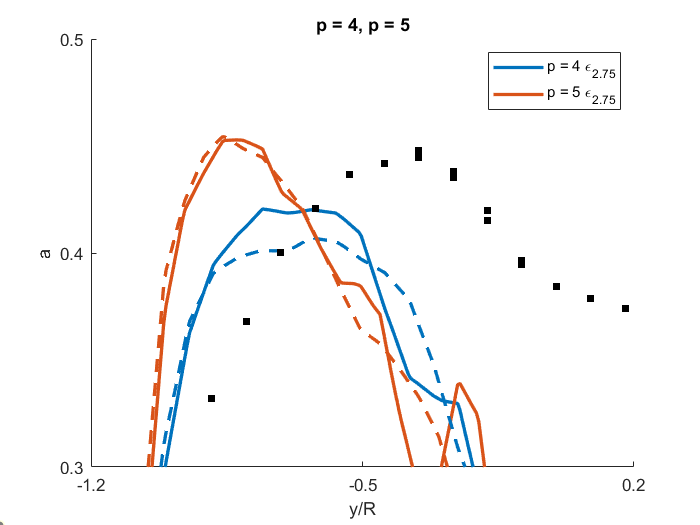}

    f)
  
\end{subfigure}
\captionsetup{justification=centering}
\caption{Velocity decay, $a$, for different polynomial order and $\epsilon_k$ parameter, $ x/D = 5$. \\  '---' AL1 '- -' AL2 '$\blacksquare$' Experimental data.  Right figures show zoomed detail of the wake peak.}
\label{fig: a vs y 3}
\end{figure}

Figures \ref{fig: a vs y 3}.a)-f) show the time-averaged velocity deficit at $x/D=5$. In this plane, we see better agreement with experiments (taking into account the missing tower and nacelle). We again observe very similar behavior of the AL1 and the AL2 implementations at low polynomials, whereas differences appear at high polynomials. Again, $\epsilon_k$ influences both implementations similarly. Overall, the AL2 method continues to provide more homogeneous profiles, while the AL1 method begins to capture oscillatory phenomena throughout the wake, which become more pronounced as we move downstream.

\begin{figure}[H]
    \centering
    \includegraphics[scale=0.4]{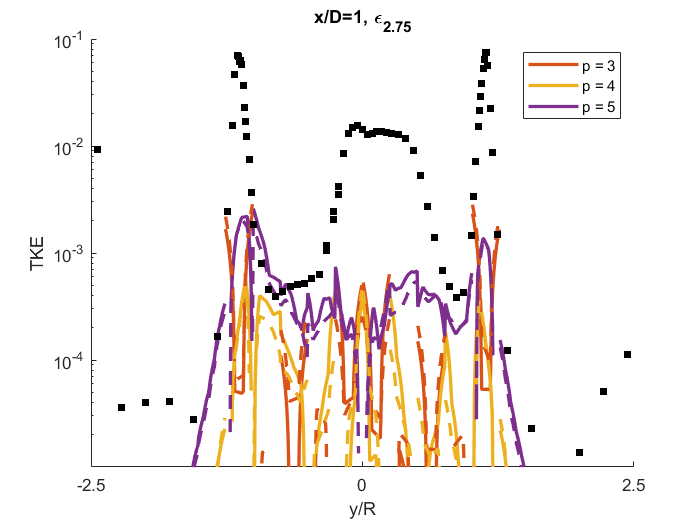}
            \includegraphics[scale=0.4]{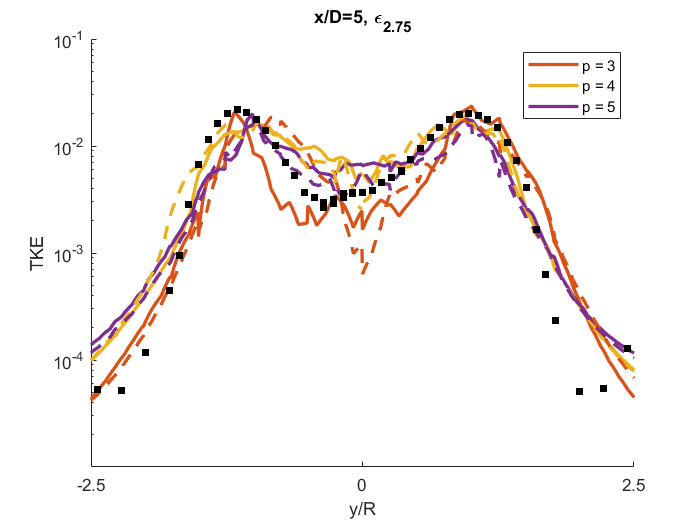}
            \captionsetup{justification=centering}
    \caption{Turbulent kinetic energy (TKE) at $x/D=1$ (left) and $x/D=5$ (right) for different polynomial order with $\epsilon_{2.75}$. \\'---' AL1, '- -' AL2, '$\blacksquare$' Experimental data.}
    \label{fig:TKE 1}
\end{figure}

\vspace{3mm}
For completeness, we show in Figure \ref{fig:TKE 1} the time averaged turbulent kinetic energy (TKE) aft the rotor plane at $x/D=1$ and 5. When analyzing the TKE, we see evidence of the need for high resolution as provided when increasing the polynomial order. Near the rotor, where the turbulent flow has not yet been fully developed, the results are distant from the experimental data. Furthermore, simulations with lower polynomial orders exhibit oscillations around values that deviate from the experimental data, confirming their inability to accurately capture the physical phenomenon (Figure \ref{fig:TKE 1}). The AL1 and the AL2 implementations provide comparable TKE distributions. 

At greater distances aft the rotor (e.g., $x/D=5$), we observe a better match with the experimental values.  This observation has already been reported \cite{Troldborg_2015} it was explained that AL methods underpredict turbulent generation but lead to accurate turbulent fluctuation in the far wake. Since the main interest of the actuator line models is to predict the wake at large enough distances from the rotor, the results obtained offer a satisfactory balance between calculation time and precision. 

\subsection{Flow structures}

\begin{figure}[]
    \centering

    \includegraphics[scale=0.5]{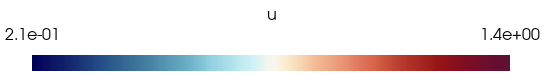}
    \vspace{2mm}
    
    \includegraphics[scale=0.40]{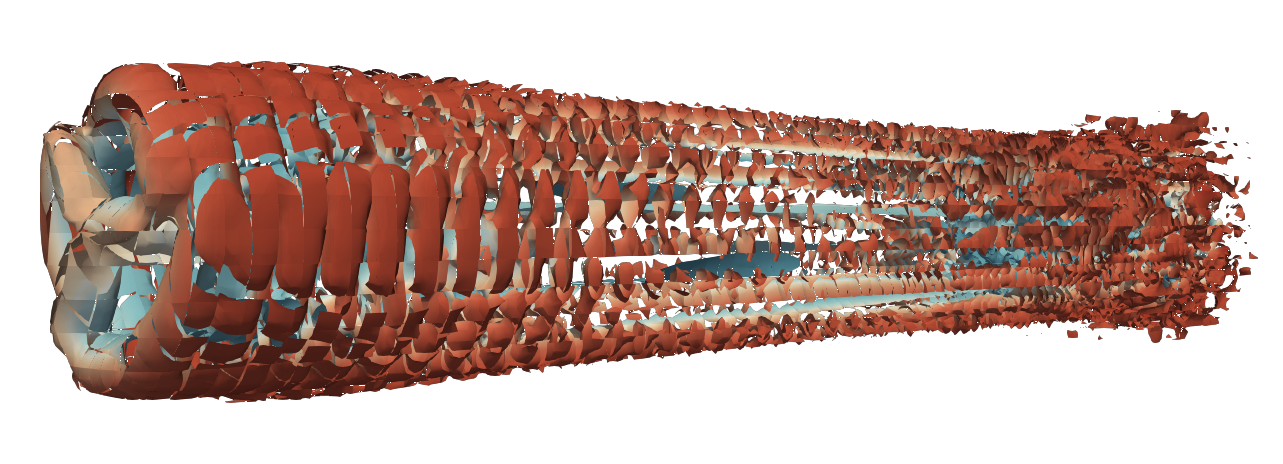} \vspace{-2mm}

    a) \vspace{-2mm}

    \includegraphics[scale=0.40]{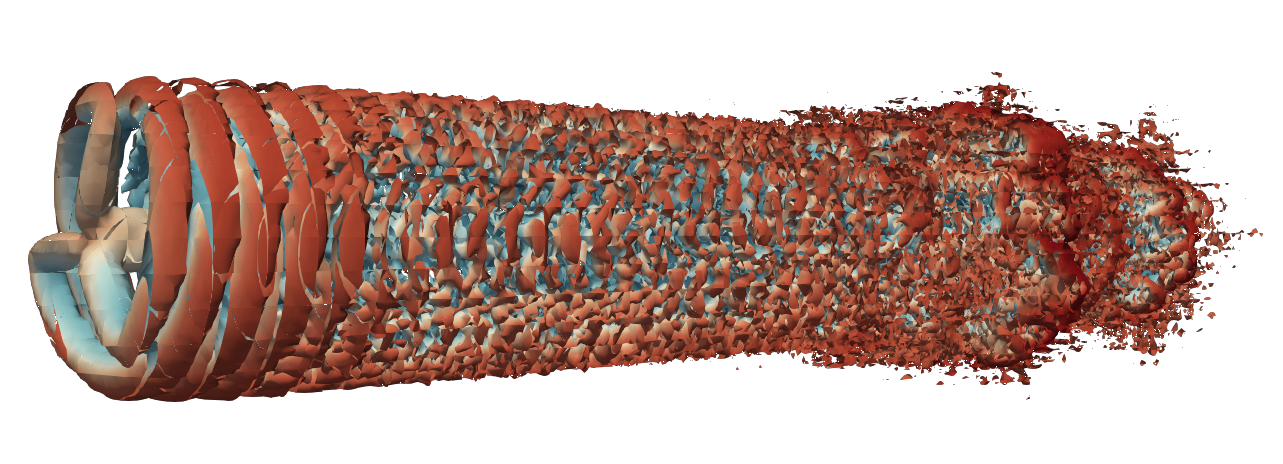} \vspace{-2mm}

    b) \vspace{-2mm}

    \includegraphics[scale=0.40]{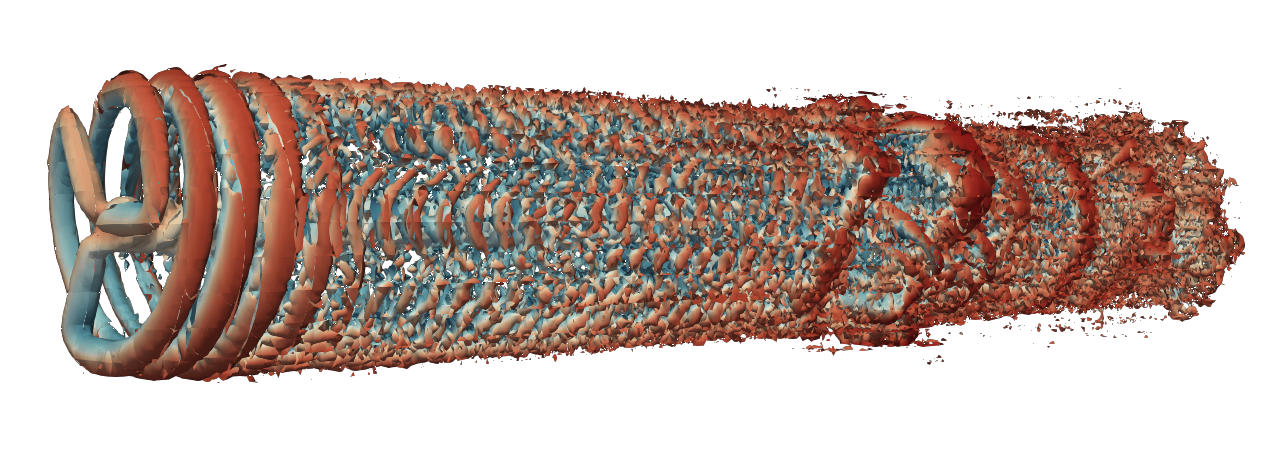} \vspace{-2mm}

    c) \vspace{2mm}

    \includegraphics[scale=0.375]{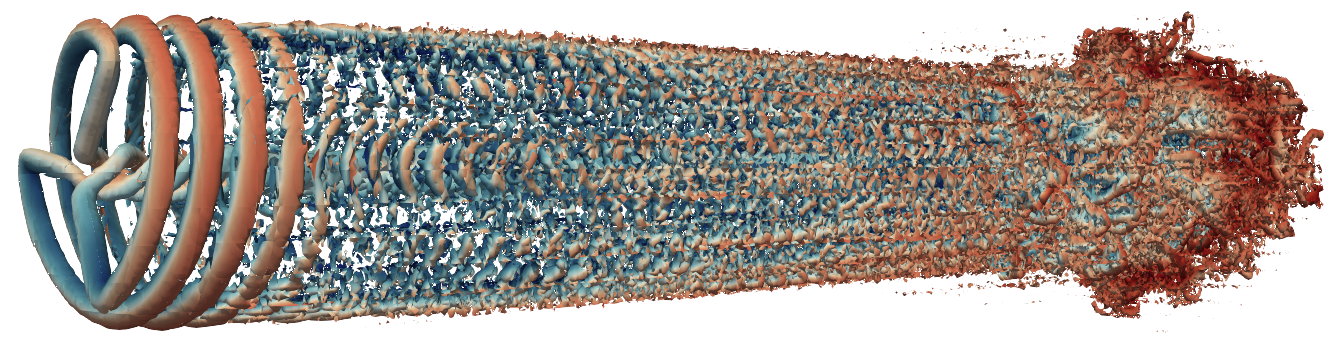} \vspace{-0mm}

    d) \vspace{-2mm}

    \vspace{-10mm}
\begin{figure}[H]
  \hspace{0mm}
  \begin{tikzpicture}
    \draw[->,-{Latex[length=2mm]}] (0,0) -- (0.67,0.05) node[right] {$x$};
    \draw[->,-{Latex[length=1.7mm]}] (0,0) -- (-0.40,0.2) node[left] {$y$};
    \draw[->,-{Latex[length=2mm]}] (0,0) -- (0,0.75) node[right] {$z$};
  \end{tikzpicture}
\end{figure}
    
     \caption{\centering{Flow structures visualized using Q-criterion iso-surfaces based on the polynomial order colored by u velocity for $\epsilon_{2.75}$ a) $p = 2$, b) $p = 3$ c) $p = 4$ and d) $p = 5$.}}
    \label{fig: Qcrit wake e275}
\end{figure}

In this final section, we show qualitatively the flow around and aft the rotor, for different polynomial orders and the two actuator line implementations. Figure \ref{fig: Qcrit wake e275} shows isocontours of the Q-criterion (colored by streamwise velocity) for polynomial orders $p=2$ to 5. A rapid improvement in resolution can be observed when increasing the polynomial order. In particular, the tip vortex becomes clear for $p > 3$ up to 1 diameter downstream of the rotor. 
Further downstream, the structures break down, enhancing mixing.

\begin{figure}[H]

\begin{subfigure}{0.5\textwidth}
  \centering
  \includegraphics[scale = 0.2]{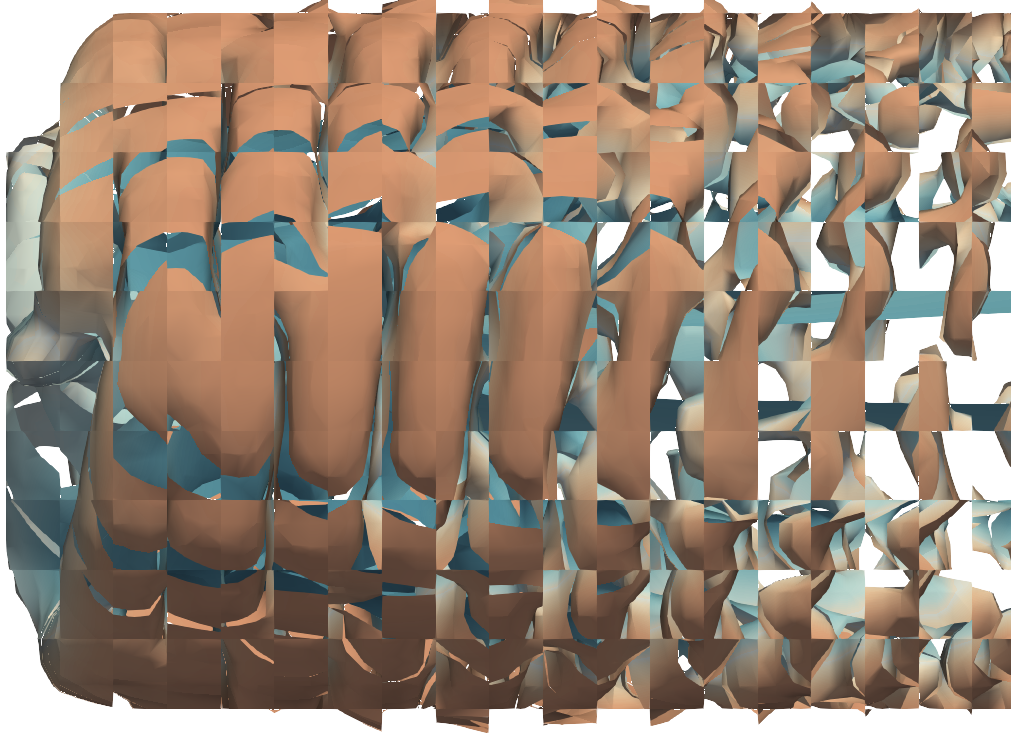}

    a)
  
\end{subfigure}
\hfill
\begin{subfigure}{0.5\textwidth}
  \centering
  \includegraphics[scale=0.2]{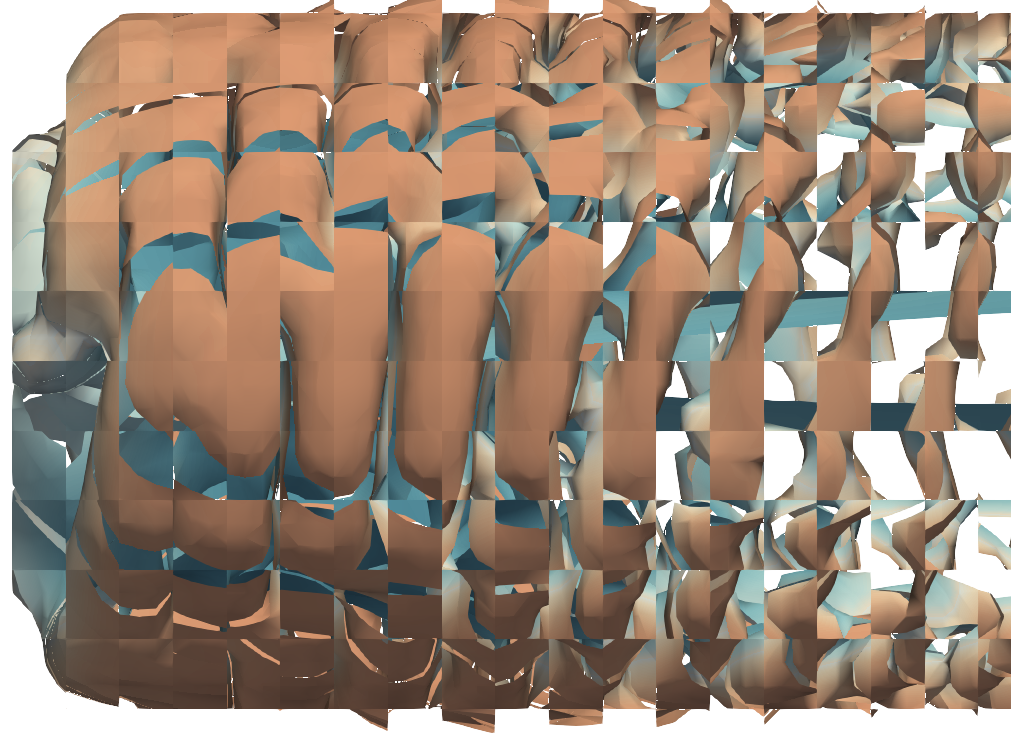}

    b)
  
\end{subfigure}

\begin{subfigure}{0.5\textwidth}
  \centering
  \includegraphics[scale = 0.2]{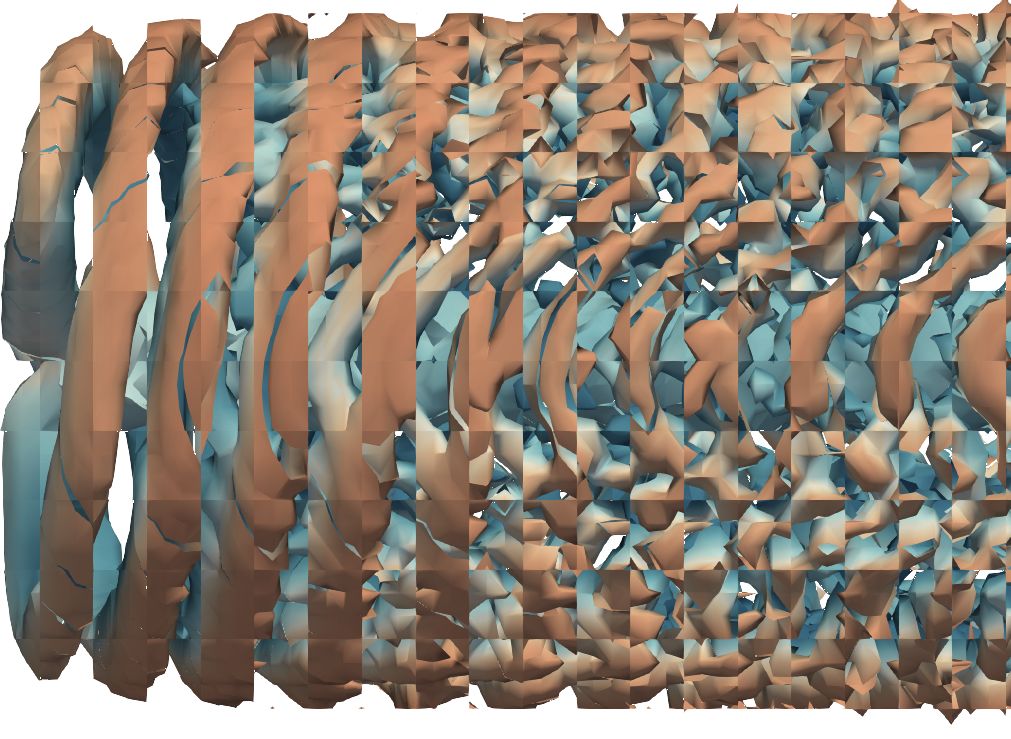}

    c)
  
\end{subfigure}
\hfill
\begin{subfigure}{0.5\textwidth}
  \centering
  \includegraphics[scale=0.2]{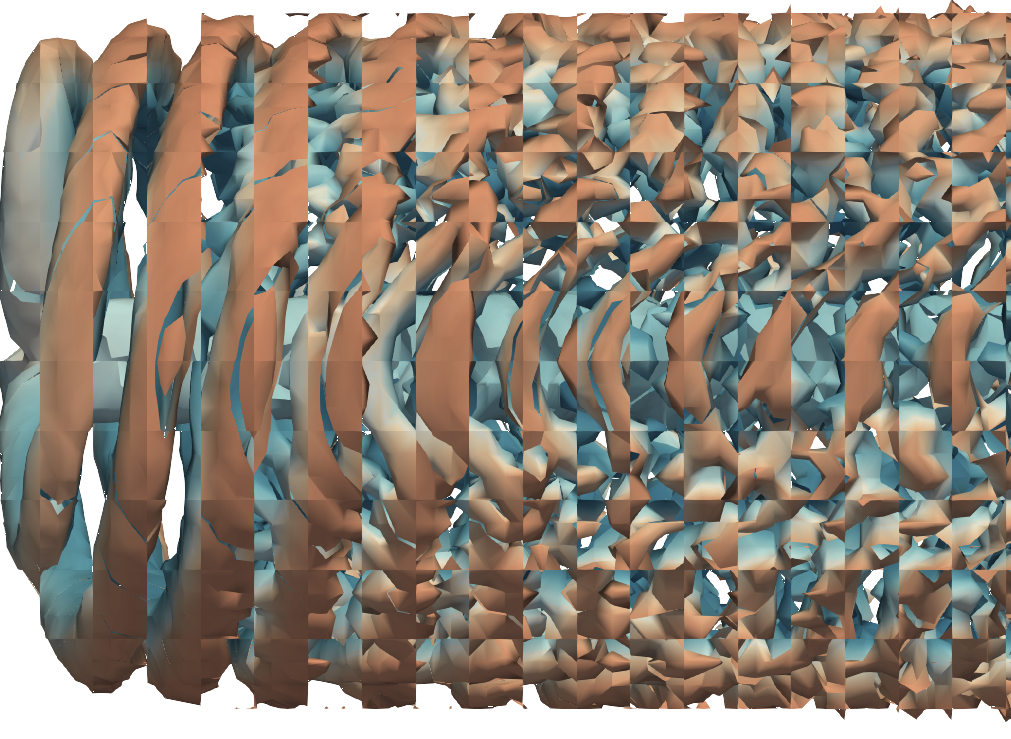}

    d)
  
\end{subfigure}

\begin{subfigure}{0.5\textwidth}
  \centering
  \includegraphics[scale = 0.2]{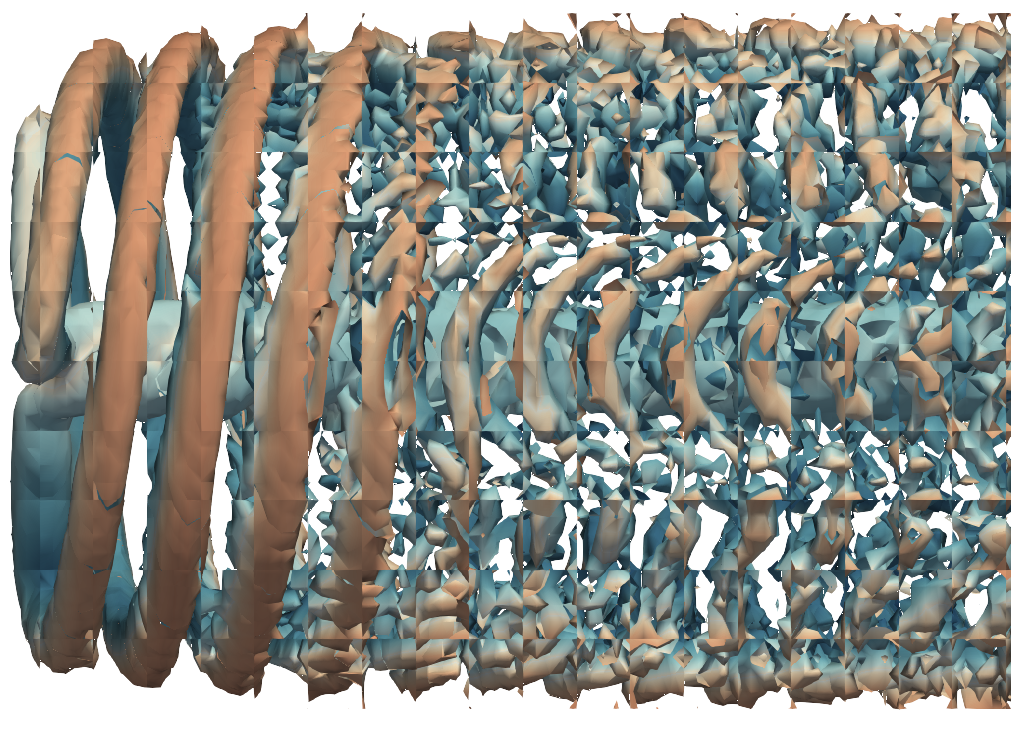}

    e)
  
\end{subfigure}
\hfill
\begin{subfigure}{0.5\textwidth}
  \centering
  \includegraphics[scale=0.2]{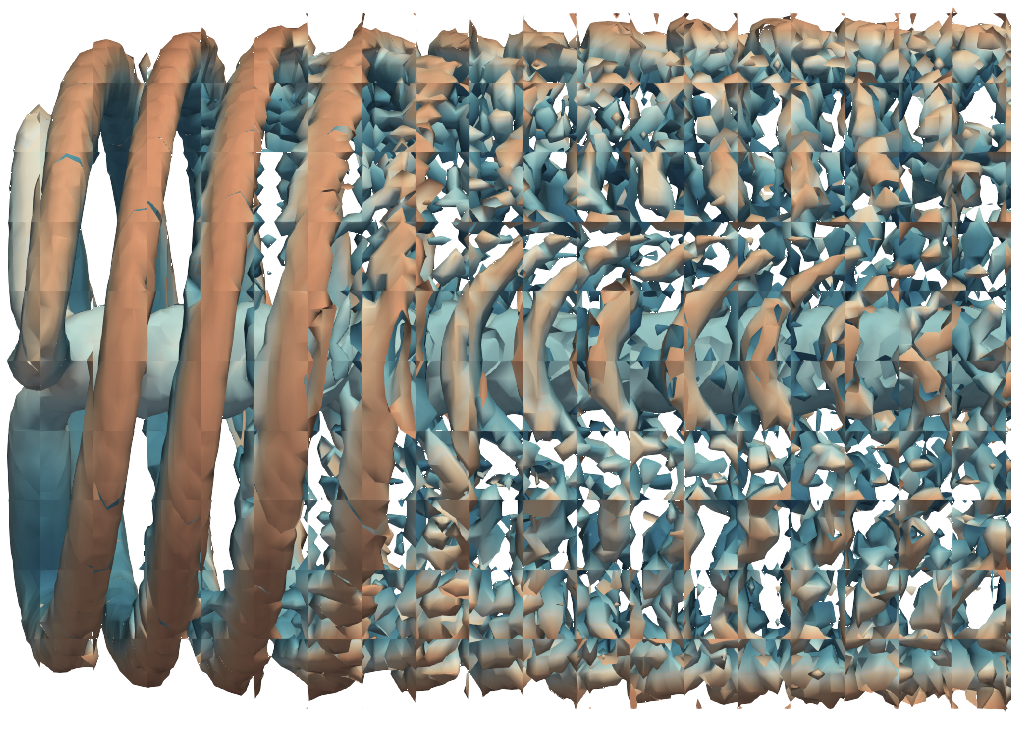}

    f)
  
\end{subfigure}
\begin{subfigure}{0.5\textwidth}
  \centering
  \includegraphics[scale = 0.2]{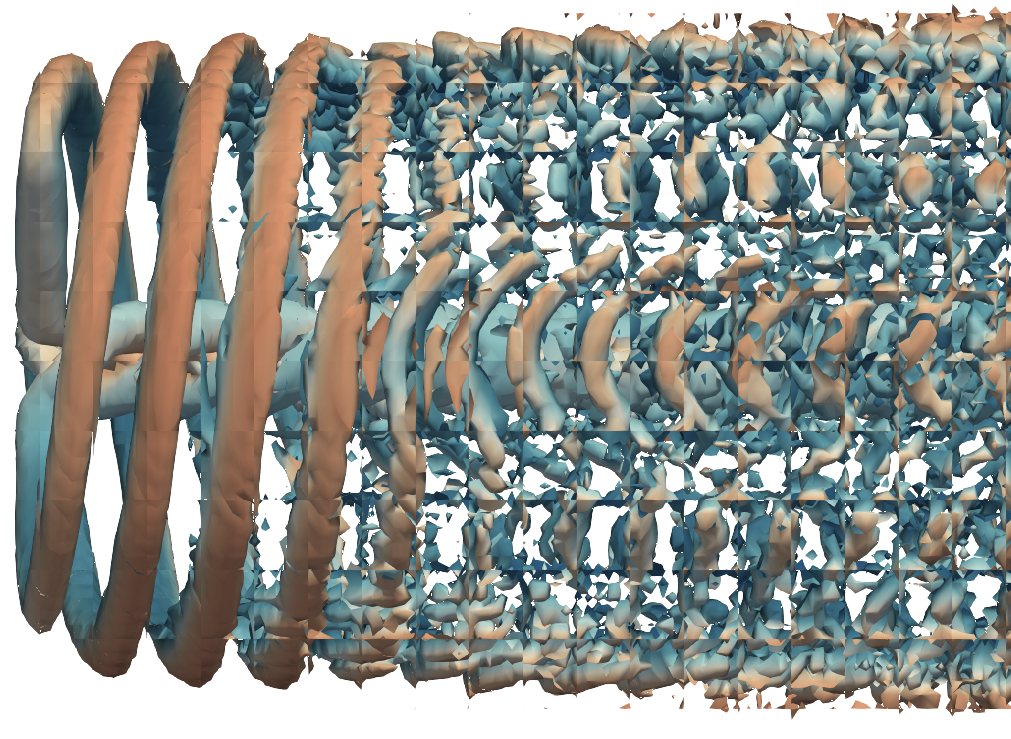}

    g)
  
\end{subfigure}
\hfill
\begin{subfigure}{0.5\textwidth}
  \centering
  \includegraphics[scale=0.2]{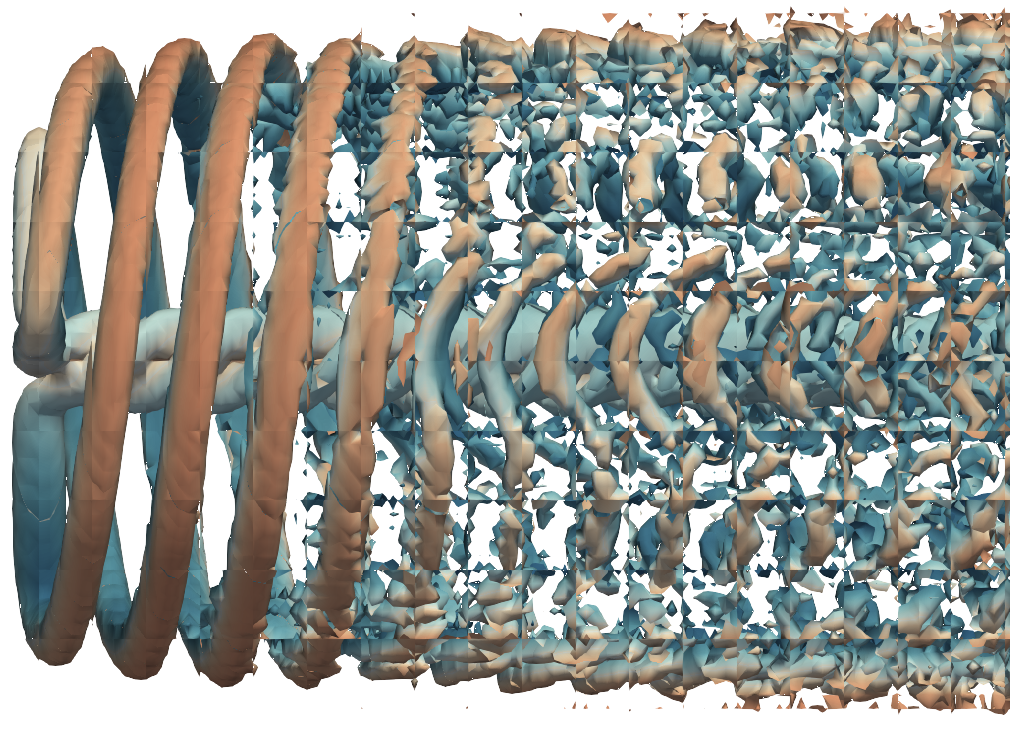}

    h)
  
\end{subfigure}

\vspace{-10mm}
\begin{tikzpicture}
    \draw[->,-{Latex[length=2mm]}] (0,0) -- (0.75,0) node[right] {$x$};
    \draw[->,-{Latex[length=2mm]}] (0,0) -- (0,0.75) node[right] {$z$};
  \end{tikzpicture}
  
\captionsetup{justification=centering}
\caption{Detail of the near wake flow structures visualized using Q-criterion isosurfaces based on the polynomial order colored by u velocity for $\epsilon_{2.75}$ a) AL1 $p = 2$, b) AL2 $p = 2$, c) AL1 $p = 3$, d) AL2 $p = 3$, e) AL1 $p = 4$, f) AL2 $p = 4$, g) AL1 $p = 5$ and h) AL2 $p = 5$.}
\label{fig: near Qcrit wake}

\end{figure}

Figure \ref{fig: near Qcrit wake} shows the near-wake Q-criterion for AL1 and AL2 and for polynomials $p=2$ to 5. Only small differences can be seen when comparing the two actuator line formulations, while as in \ref{fig: Qcrit wake e275}, there is a clear improved resolution when increasing the polynomial order.

\begin{figure}[H]
    \centering

    \includegraphics[scale=0.5]{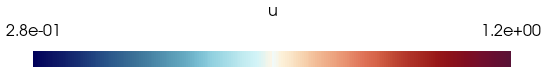}
    \vspace{2mm}
    
    \includegraphics[scale=0.405]{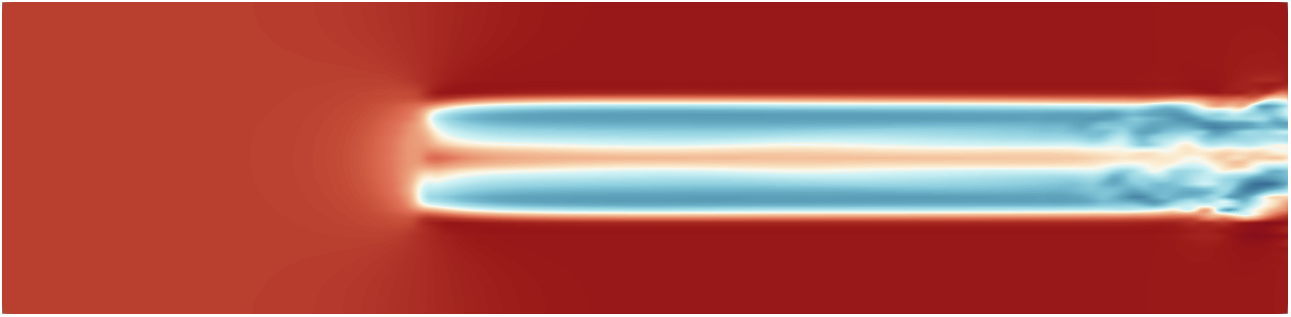}

    a)
    
    \includegraphics[scale=0.417]{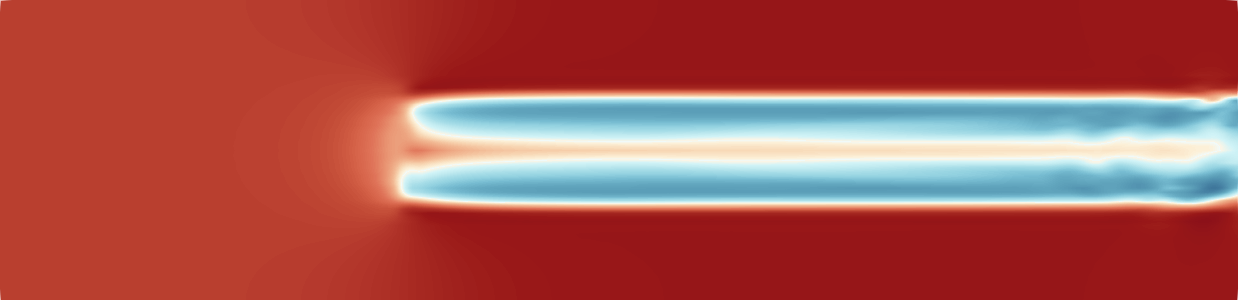}
   
    b)
    
    \includegraphics[scale=0.408]{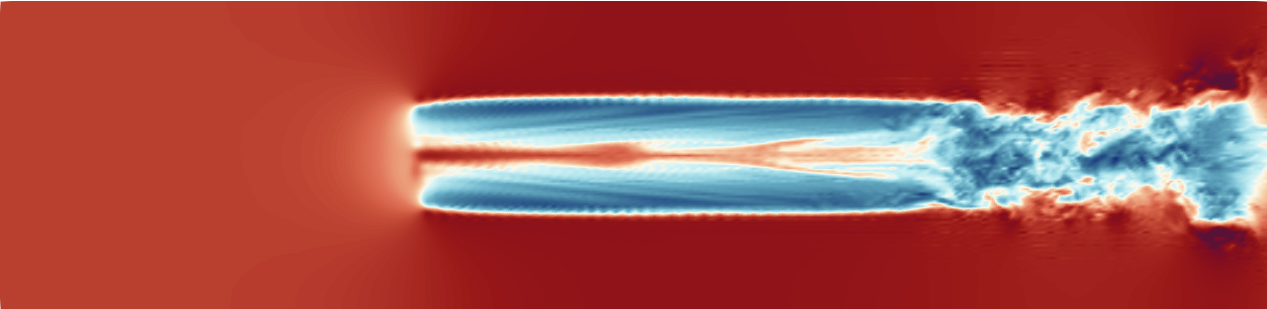}
    
    c)
    
    \includegraphics[scale=0.417]{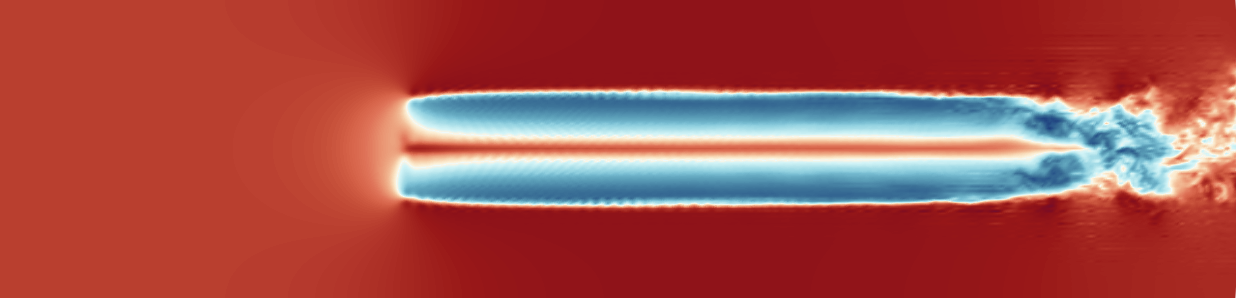}
    
    d)

\vspace{-19.5mm}
\begin{figure}[H]
  \hspace{5.3mm}
  \begin{tikzpicture}
    \draw[->,-{Latex[length=2mm]}] (0,0) -- (0.75,0) node[below] {$x$};
    \draw[->,-{Latex[length=2mm]}] (0,0) -- (0,0.75) node[left] {$y$};
  \end{tikzpicture}
\end{figure}

\captionsetup{justification=centering}
    \caption{Section of the instantaneous velocity field along the $x$-axis for $\epsilon_{2.75}$, a) AL1 $p = 2$, b) AL2 $p = 2$, c) AL1 $p = 5$ and d) AL2 $p = 5$. Al of them with $\epsilon_{2.75}$}
    \label{fig: wake e275}
    
\end{figure}

Finally, we show the instantaneous streamwise velocity at a spanwise plane only for polynomials of order 2 and 5, when using the AL1 and the AL2 implementations. Using $p=2$, the flow fields look very similar, but when using $p=5$, the flow fields look substantially different. Indeed, for the AL1 formulation, oscillations appear near the rotor center (see Figure \ref{fig: wake e275}.c), which are not present for the AL2 formulation (see Figure \ref{fig: wake e275}.d)). 

\begin{figure}[H]
      \centering
      \includegraphics[scale = 0.4]{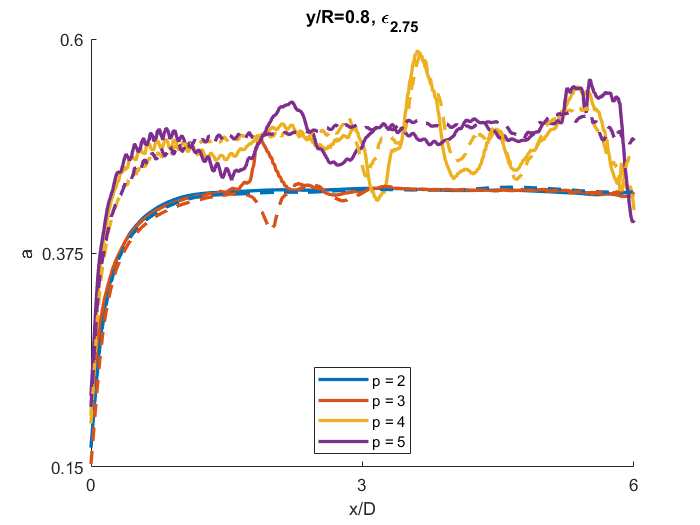}   
      \includegraphics[scale = 0.4]{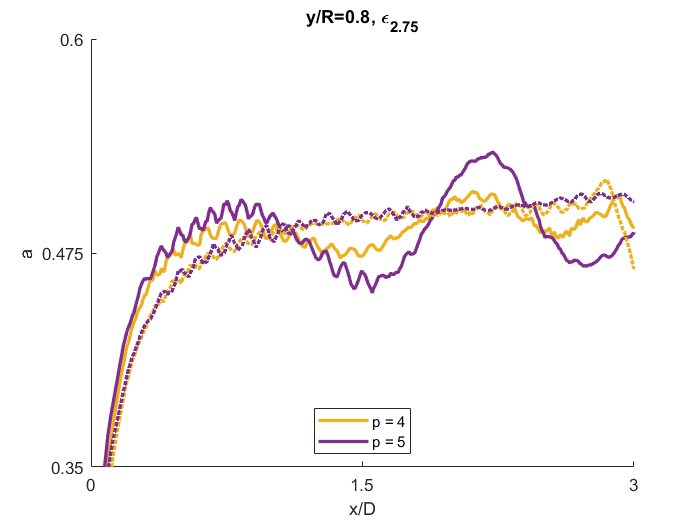}   
      \captionsetup{justification=centering}
      \caption{Stream-wise velocity deficit (a) for different polynomial orders ($p=2$ to 5) with a fixed $\epsilon_{2.75}$.\\ '---' AL1 '- -' AL2}
      \label{fig:a vs x}
  \end{figure}

\begin{figure}[H]

  \centering
  \includegraphics[trim={16cm 9cm 18cm 9cm}, clip,width=0.49\linewidth]{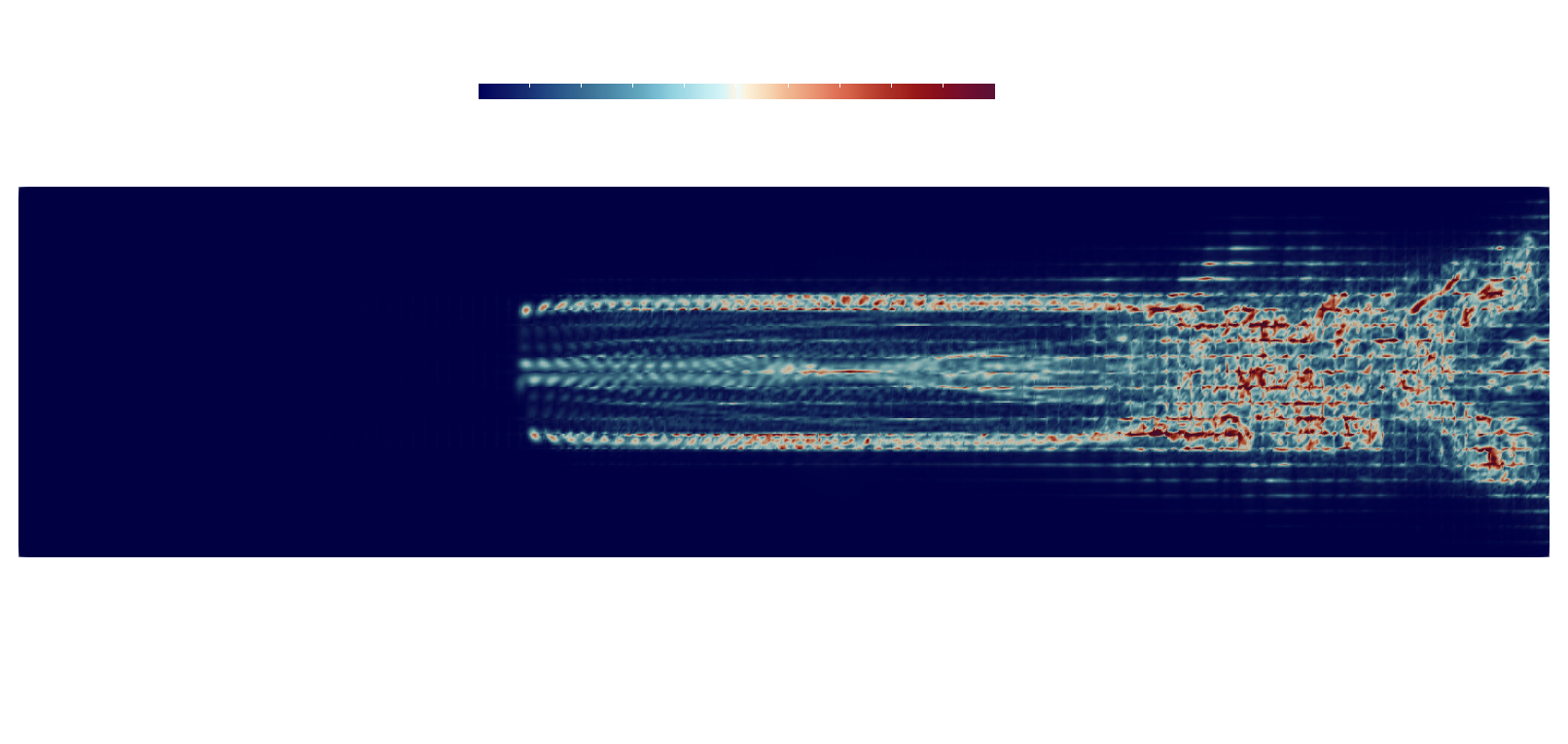}
  \includegraphics[trim={16cm 9cm 18cm 9cm}, clip,width=0.49\linewidth]{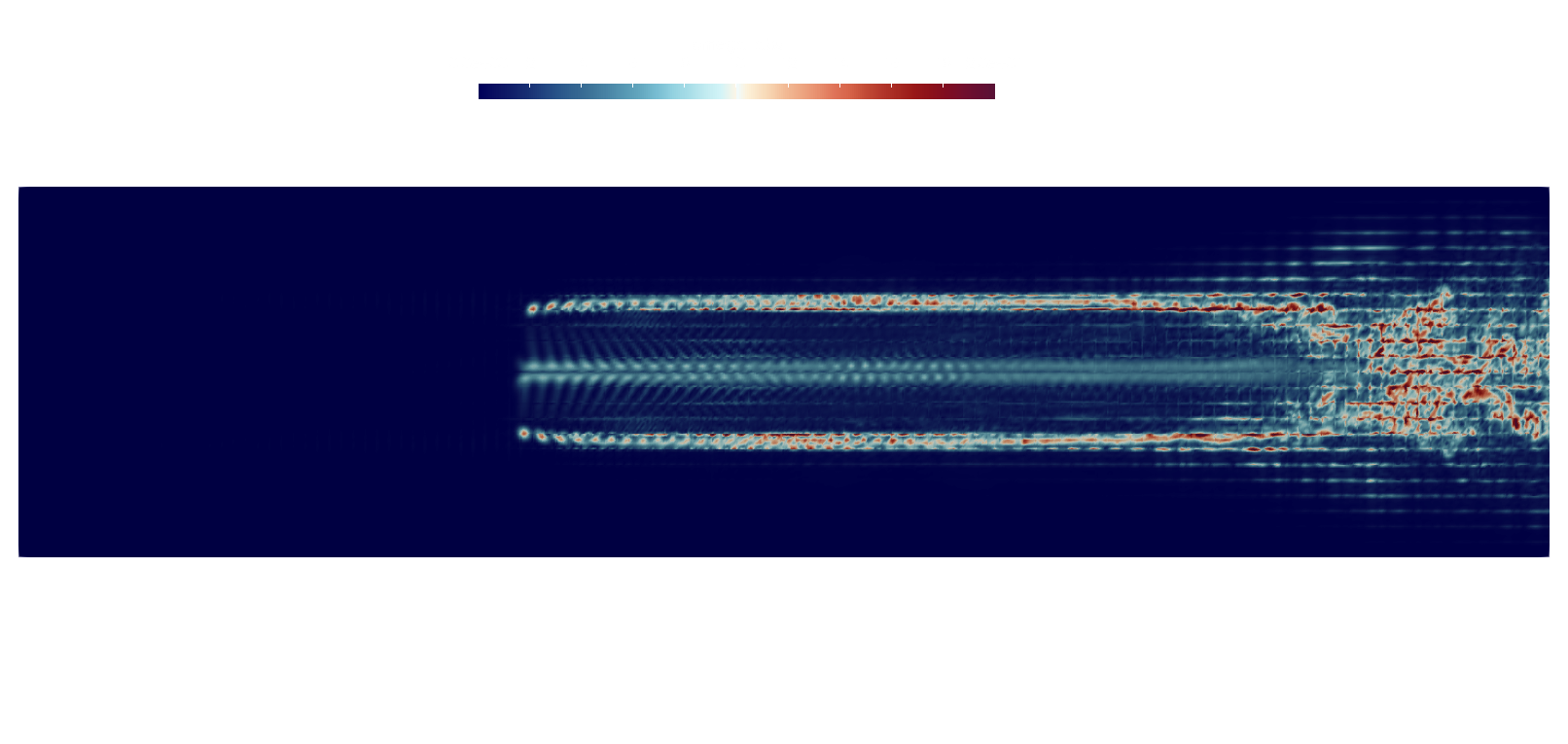}
 
\captionsetup{justification=centering}
\caption{Instantaneous vorticity fields along the x/y-plane for polynomial order $p=5$ and $\epsilon_{2.75}$: Cell average method, AL1 (left) and projection method, AL2 (right).}
\label{fig:vorticity}

\end{figure}

Guided by the last intriguing foundlings, where we found oscillations appeared in the AL1 method (and not present in the AL2). We depict in \ref{fig:a vs x} the evolution of the velocity along the stream-wise direction at a fixed height corresponding to 80\% of the rotor radius and for various polynomial orders. We observe that for low polynomials both AL1 and AL2 are identical, while for high polynomial order differences are noticeable. To clarify the existence of long spatial oscillations for AL1 (and not present in AL2) we show the instantaneous flow fields of vorticity magnitude in figure \ref{fig:vorticity}.
The figure shows that the AL1 formulation generates long spatial oscillations near the root that are not present in the AL2 case. When checking other numerical schemes and actuator line formulations, see, for example, \cite{Oxford_sim}, we do not observe oscillations near the blade root. We conclude that for low polynomial orders, both AL1 and AL2 implementations provide identical results when checking average results and instantaneous velocity fields. However, when increasing the polynomial, we favor the AL2 formulation, since AL1 can induce non-physical oscillations in the solution.

These nonphysical oscillations are attributed to the use of cell averaging (AL1) when using high polynomials (here $p=5$), which introduces errors when dealing with large gradients inside the h-mesh. 
For low polynomial orders ($p\leq3$) we see the same behavior for AL1 and AL2, and it is only when the polynomial increases beyond 3 that we observe nonphysical oscillations for AL1.
Finally, when reexamining the time-averaged results (see, for example, figure \ref{fig: a vs y 1}.f)), we observe the footprint of the long spatial oscillation in the averages of the velocity field in the wake (only for $p=4$ and 5).
We conclude that the proposed actuator line implementations AL1 (averaging) and AL2 provide almost identical results for low polynomial orders, but the implementation of AL2 (projection) is recommended when using high polynomial orders.

\section{Conclusions}\label{sec:conclusion}

In this work, we have demonstrated the effectiveness of integrating the actuator line method with high-order $h/p$ solvers (order ranging from 3rd to 6th) for wind turbine modelling. By representing the rotating turbine blades as distributed force lines, the actuator line method captures the blade's influence on the flow field in a computationally efficient manner. Through extensive numerical simulations, we have shown that the combined approach of actuator lines and high-order $h/p$ solvers offers flexibility to improve the flow solution by means of increasing the polynomial order (without needing to remesh). Furthermore, the following conclusions have been obtained.

\begin{itemize}
\item Near and far wake first order statistics (e.g., velocity) are well captured once a sufficient polynomial order is selected. Second-order statistics (e.g., turbulent kinetic energy) are well-captured in the far wake.
\item High polynomial orders prove advantageous in capturing the correct physics.
\item The high-order version of the smoothing Gaussian: $\epsilon_k=k\times\Delta_{grid} = k\times\frac{(\Delta_x \Delta_y \Delta_z)^{\frac{1}{3}}}{p+1}$ with $k\in [1, 4]$ provides consistent results for the two actuator line implementations tested. Reducing $k$ leads to sharper gradients in the flow.
\item The two implemented AL methods (averaging and projection) showed similar accuracy and robustness when comparing forces along blades and wake averages. 
\item The AL1 formulation (averaging) is less costly for high polynomial orders than AL2 (projection).
\item The AL1 (averaging) and AL2 (projection) formulations provide very similar results for low polynomial orders (equivalent to low order methods such as finite volume or finite elements), but differ for high order polynomials ($p\geq4$). In these cases, the AL1 implementation introduces non-physical oscillations near the rotor center, whose footprint is visible within the turbine wake (in the instantaneous and averaged flow).
\end{itemize}

As shown, when integrating actuator lines, often conceived for low-order methods, into high-order $h/p$ solvers, caution is needed to avoid nonphysical flows. With correct blade modelling, high-order $h/p$ solvers offer promising prospects for conducting large-scale wind turbine simulations of farms.

\section*{Acknowledgments}
Esteban Ferrer, Oscar Mariño and Ananth Sivaramakrishnan would like to thank the support of the
Agencia Estatal de Investigación (for the grant "Europa Excelencia 2022" Proyecto EUR2022-134041/AEI/10.13039/501100011033) y del Mecanismo de Recuperación y Resiliencia de la Unión Europea.  
Esteban Ferrer also acknowledges the help of the Comunidad de Madrid and Universidad Politécnica de Madrid for the Young Investigators award: APOYO-JOVENES-21-53NYUB-19-RRX1A0. 
This research has received funding from the European Union (ERC, Off-coustics, project number 101086075). Views and opinions expressed are, however, those of the authors only and do not necessarily reflect those of the European Union or the European Research Council. Neither the European Union nor the granting authority can be held responsible for them.
Stefano Colombo thanks the European Union Horizon 2020 Research and Innovation Program under the Marie Sklodowska-Curie grant agreement No 955923 for the Ssecoid project.
Finally, all authors gratefully acknowledge the Universidad Politécnica de Madrid (www.upm.es) for providing computing resources on the Magerit Supercomputer.

\section*{Authors’ contributions}
\begin{itemize}
    \item Oscar A. Marino: Conceptualization, Data curation, Formal analysis, Investigation, Methodology, Software, Validation, Visualization, Writing - original draft.
    \item Raúl Sanz: Conceptualization, Data curation, Formal analysis, Investigation, Software, Validation, Visualization, Writing - original draft.
    \item Stefano Colombo: Conceptualization, Formal analysis, Software, Validation, Visualization, Writing - original draft.
    \item Ananth Sivaramakrishnan: Formal analysis, Investigation, Software, Validation, Visualization, Writing - original draft.
    \item Esteban Ferrer: Conceptualization, Methodology, Validation, Funding acquisition,  Project administration, Resources, Supervision, Writing - review \& editing.
\end{itemize}

\section*{Competing interests}
The authors have no relevant financial or non-financial interests to disclose.

\section*{Data availability}
The data that support the findings of this study are available from the corresponding author, Oscar A. Marino, upon reasonable request. Furthermore, the CFD code used to simulate the wind turbines is available in the open source code HORSES3D~\cite{HORSES_3D}.

\appendix
\section{Compressible Navier-Stokes solver}
\label{sec:cNS}
 In this work we solve the 3D Navier-Stokes equations for laminar cases and supplement the equations with the Vreman LES model for turbulent flows. The 3D Navier-Stokes equations when including the  Vreman model can be compactly written as:
%
\begin{equation}
\boldsymbol{q}_t+ \nabla \cdot {\ssvec{F}}_e = \nabla\cdot\ssvec{F}_{v,turb},
\label{eq:compressibleNScompact}
\end{equation}
where $\boldsymbol{q}$ is the state vector of large scale resolved conservative variables $\boldsymbol{u} = [ \rho , \rho v_1 , \rho v_2 , \rho v_3 , \rho e]^T$, $\ssvec{F}_e$ are the inviscid, or Euler fluxes,
\begin{equation}
\ssvec{F}_e = \left[\begin{array}{ccc} \rho v_1 & \rho v_2 & \rho u_3 \\
                                                                                \rho v_1^2 + p & \rho v_1v_2 & \rho v_1v_3 \\
                                                                                	\rho v_1v_2 & \rho v_2^2 + p & \rho v_2v_3 \\
                                                                                	\rho v_1v_3 & \rho v_2v_3 & \rho v_3^2 + p \\
                                                                                	\rho v_1 H & \rho v_2 H & \rho v_3 H
\end{array}\right],
\end{equation}
where $\rho$, $e$, $H=E+p/\rho$, and $p$ are the large scale density, total energy, total enthalpy and pressure, respectively, and $\vec{v}=[v_1,v_2,v_3]^T$ is the large scale resolved velocity components. Additionally, $\ssvec{F}_{v,turb}$ defines the viscous and turbulent fluxes,
\begin{equation}
\ssvec{F}_{v,turb}= \left[\begin{array}{ccc}0 & 0 & 0\\
 \tau_{xx} & \tau_{xy} & \tau_{xz} \\
 \tau_{yx} & \tau_{yy} & \tau_{yz} \\
 \tau_{zx} & \tau_{zy} & \tau_{zz} \\
 \sum_{j=1}^3 v_j\tau_{1j} + \kappa T_x& \sum_{j=1}^3 v_j\tau_{2j} + \kappa T_y& \sum_{j=1}^3 v_j\tau_{3j} + \kappa T_z
\end{array}\right],
\label{eq:viscousfluxes}
\end{equation}
where $\kappa$ is the thermal conductivity, $T_x, T_y$ and $T_z$ denote the temperature gradients and the stress tensor $\boldsymbol{\tau}$ is defined as $\boldsymbol{\tau} = (\mu+\mu_t)(\nabla \vec{v} + (\nabla \vec{v})^T) - 2/3(\mu+\mu_t) \boldsymbol{I}\nabla\cdot\vec{v}$, with $\mu$ the dynamic viscosity, $\mu_t$ the turbulent viscosity (in this work defined through the Vreman 
model) and $\boldsymbol{I}$ the three-dimensional identity matrix. Note that when solving laminar flows, it suffices to set $\mu_t=0$ and reinterpret the large-scale resolved components as the only components (there are no under-resolved components).
The dynamic turbulent viscosity using the Vreman \cite{Vreman_2004} model is given by: 
\begin{equation}
\begin{split}
    &\mu_t = C_v \rho\sqrt{\frac{B_\beta}{\alpha_{ij}\alpha_{ij}}},\\
    &\alpha_{ij} = \frac{\partial u_j}{\partial x_i},\\
    &\beta_{ij} = \Delta^2\alpha_{mi}\alpha_{mj},\\
    &B_\beta = \beta_{11}\beta_{22} -\beta_{12}^2 +\beta_{11}\beta_{33} -\beta_{13}^2 +\beta_{22}\beta_{33} -\beta_{23}^2,
\end{split}
\label{eq-iLES:LES_vreman}
\end{equation}

\noindent where $C_v=0.07$ is the constant of the model. 
The Vreman LES model adjusts the model parameters based on the local flow characteristics and automatically reduces the turbulent viscosity in laminar, transitional, and near-wall regions, allowing to capture the correct physics. 

\bibliographystyle{elsarticle-num-names} 
\bibliography{refs}

\end{document}